\documentclass[aps,amssymb,amsmath,superscriptaddress,prl,reprint,longbibliography]{revtex4-2}

\usepackage{graphicx}
\usepackage{color}
\usepackage{amsmath}
\newcommand{\stkout}[1]{\ifmmode\text{\sout{\ensuremath{#1}}}\else\sout{#1}\fi}
\usepackage{enumitem}
\usepackage{amssymb}
\usepackage{hyperref}
\usepackage{cancel}
\usepackage{ulem}
\usepackage{multirow}
\usepackage{bm}
\usepackage{physics}
\usepackage{pifont}

\newcommand{\be}{\begin{equation}}
	\newcommand{\ee}{\end{equation}}
\newcommand{\bea}{\begin{eqnarray}}
	\newcommand{\eea}{\end{eqnarray}}
\renewcommand\vec[1]{\ensuremath\mathbf{#1}} 

\usepackage{amsfonts, relsize, color}
\usepackage{graphicx}
\usepackage{color}
\usepackage{comment}

\usepackage{xcolor}
\hypersetup{
	colorlinks,
	linkcolor={red!50!black},
	citecolor={green!50!black},
	urlcolor={blue!50!black}
}
\usepackage{booktabs} 
\begin{document}

\title{Valley polarization of chiral excitonic bound states induced by band geometry}

\author{Archisman Panigrahi}\email{archi137@mit.edu}\affiliation{Department of Physics, Massachusetts Institute of Technology, Cambridge, MA 02139, USA}
\author{Daniel Kaplan}\email{d.kaplan1@rutgers.edu}\affiliation{
Center for Materials Theory, Department of Physics and Astronomy, Rutgers University, Piscataway, NJ 08854, USA}
\affiliation{Department of Physics, University of Central Florida, Orlando, FL 32816, USA}
\affiliation{NanoScience Technology Center, University of Central Florida, Orlando, FL 32816, USA}

\begin{abstract}
Van der Waals (vdW) materials provide a rich platform for exploring the interplay of interactions, topology, and paired-electron phases. We study how the Berry phase reshapes excitonic pairing in a double-well dispersion representative of layered vdW systems. By computing the temperature-versus-Berry flux phase diagram of the system, we find parameter ranges where finite angular momentum excitons are favored, including chiral states. Strikingly, the condensed angular momentum channel evolves with Berry flux, revealing a pairing problem with no analogue in a hydrogen atom in a uniform magnetic field, where angular momentum states never cross. We then turn to a model of multilayer rhombohedral graphene and examine the effects of trigonal warping. Once continuous rotational symmetry is broken, excitons mix multiple angular momenta, and for a range of parameters we find a variety of linear combination of angular momenta ($s, p, f$ and $g$) in the ground state. {Due to bosonic interactions, the ground state is spontaneously valley polarized for low exciton densities}. Our results point to the possibility of chiral excitonic condensates, and spontaneous symmetry breaking through many-body condensation.
\end{abstract}
\date{\today}

\maketitle

\paragraph*{Introduction ---}
Bloch band geometry and electronic topology have been shown to have important effects on quantum material properties, from transport to collective phenomena \cite{Sundaram_wave-packet_1999,Xiao-Chang-Niu_Berry,Vanderbilt_berry_2018,Peotta_superfluidity_2015,Julku_geometric_2016,Dillenschneider_2008,Liang_band_2017,Kwan_excitonic_2022,Guerci2025}. 
When interactions are considered, the projection of the Coulomb repulsion onto low-energy bands endows the interaction vertex with geometric form factors, so that Berry phases can reshape the different channels that are mediated by the interaction. This was shown to apply to 
pumped excitonic states \cite{Yao_exciton_2008,Wu_exciton_2015,Srivastava_signatures_2015,Zhou_Berry_2015,
Xie_topological_2024,Guerci2024,Martinez2025, Zeng_topological_excitation, zverevich2026spintripletpairedwignercrystal, Hao_neutral_2017, Steinhoff_biexciton_2018, Selig_ultrafast_2019, Feng_highly_2024}. While the normal state of many systems with topological bands is achiral (e.g., due to time-reversal symmetry, TRS), it is natural to wonder whether a globally achiral normal state can nevertheless condense into a particle-hole paired phase with spontaneous chirality and broken time-reversal symmetry.
This question is especially timely given that layered graphene, transition-metal dichalcogenides (TMD), such as monolayer WTe$_2$ or stacked heterobilayers, and moir\'e van der Waals heterostructures now show evidence for excitonic insulating or condensate behavior \cite{Nandi_drag_2012,Ju_tunable_2017,
wang2019evidence,ma2021strongly,sun2022evidence,chen2022excitonic,zhang2022correlated,liu2022crossover,qi2023thermodynamic, Nguyen_quantum_2025, Qi_observation_2026}, in close proximity to symmetry broken phases. From a theoretical perspective, the defining feature of these materials is that they host low-energy bands with tunable quantum geometry \cite{Koshino_trigonal_2009,Kumar_flat_2013} (by extension, Berry curvature) making them natural platforms in which to ask how quantum geometry reorganizes the excitonic pairing problem under a realistic interaction \cite{Hu_quantum_metric_2022}. Although spin and valley polarized states in excitonic condensates have been discussed in toy models \cite{Fernandez_1997,Wu_Macdonald2015}, the role of quantum geometry and interactions in generating spontaneous chirality has not been explored.

\begin{figure}[t!]
    \centering
    \includegraphics[width=0.99\linewidth]{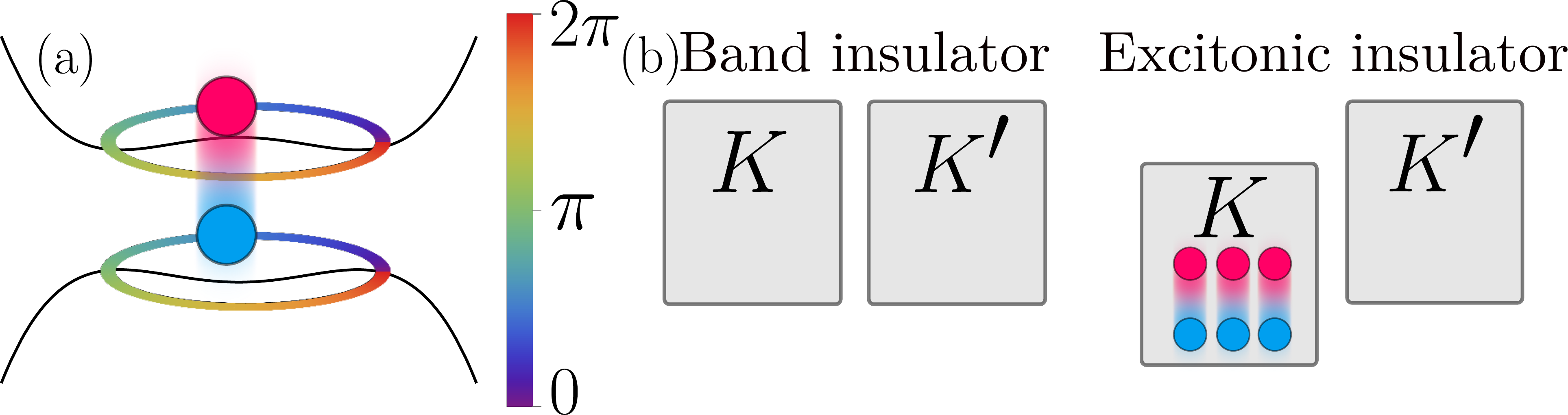}
    \caption{(a) A schematic of a chiral $p$-wave exciton in the Mexican hat dispersion model. The colorbar displays its complex phase winding. (b) Illustration of spontaneous valley polarization of excitons stabilized by a generalized Stoner mechanism.}
    \label{fig:mexican_hat_cartoon}
\end{figure}

Here we revisit the excitonic pairing problem at the mean-field level by extending the classic treatment of J\'erome \textit{et al.} \cite{Keldysh_instability_1965,Jerome_Excitonic_1967,Kohn_exciton_1967,Halperin_Rice_1968}. We include Bloch-state form factors \cite{Quintela_tunable_2022, Davenport_berry_2026, hu2022quantum, Scammell_2023, Quintela_theoretical_2022} which manifest non-trivial quantum geometry \cite{yu2025quantum}. Within a single-exciton ansatz and a generic double-well (“sombrero”) dispersion relevant to layered van der Waals systems, we show, both in the exciton spectrum and in the BCS-like gap equations, that Berry flux stabilizes finite-angular-momentum excitons and can then switch the leading pairing channel to be odd. This behavior has no analog in the corresponding two-dimensional hydrogenic problem in a uniform magnetic field, where we find no comparable even-to-odd level crossing. We then turn to rhombohedral multilayer graphene, where trigonal warping reduces continuous rotational symmetry and mixes angular-momentum channels, enabling mixed-symmetry chiral states \cite{Koshino_trigonal_2009,Kumar_flat_2013,Dong_anomalous_2024,Soejima_anomalous_2024,Tan_parent_2024}. Even in this setting, states with odd components which energetically favored emerge over a broad range of interaction strengths, displacement fields, and dielectric environments. This occurs due to a mixture of the inherent Berry phase of the bands, the trigonal warping, and the rhombohedral structure.
Finally, we argue that in multivalley systems exchange in the multi-exciton problem favors spontaneous valley polarization, in close analogy with a Stoner instability.
\begin{figure}[t!]
    \centering
    \includegraphics[width=0.99\linewidth]{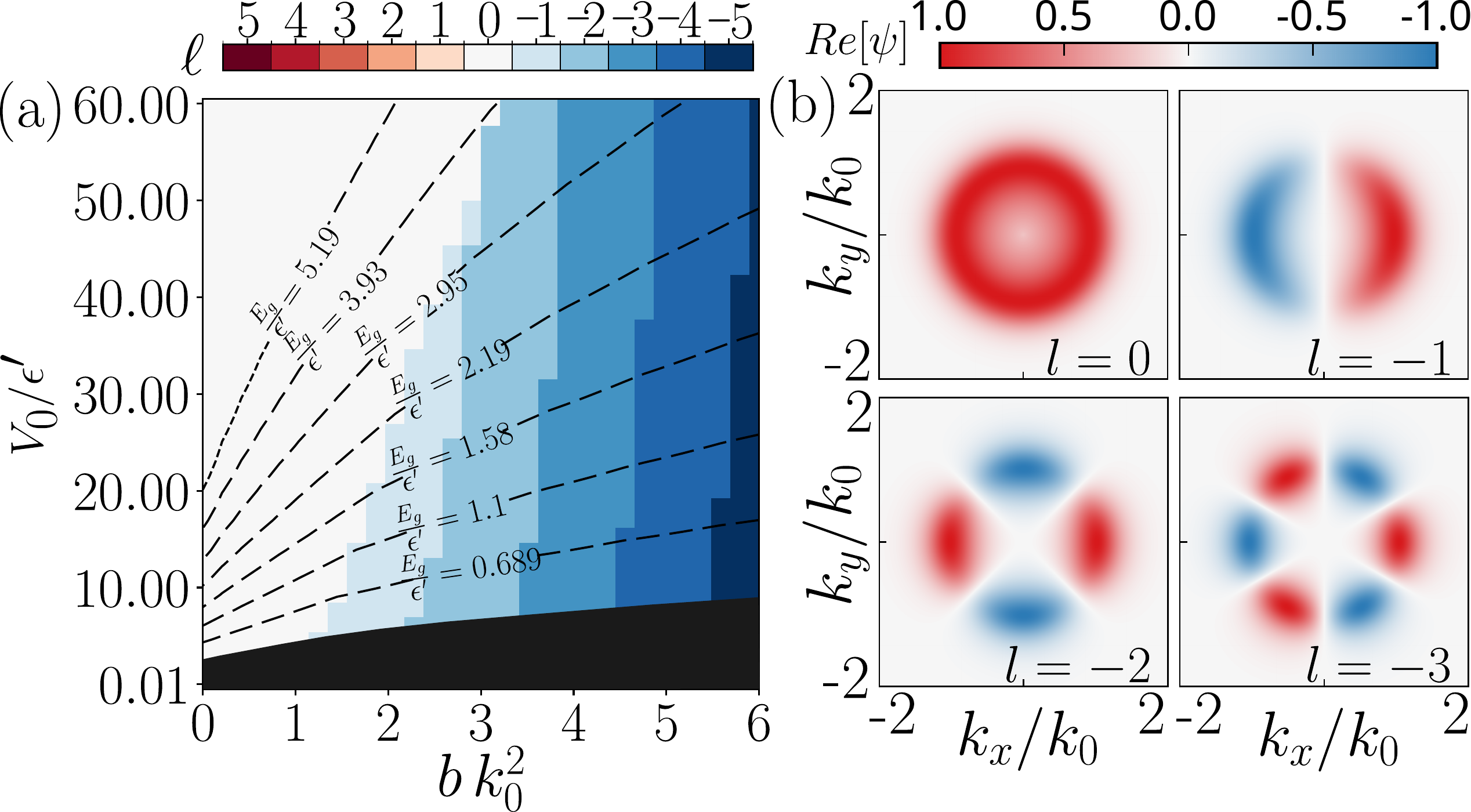}
    \caption{(a) Phase diagram for a single exciton in the Mexican hat toy model (Eq.~\eqref{eq:bethe-salpeter-single}) for gate distance $d_{\rm gate} =10/k_0$ and bandgap $E_g/\epsilon' = 0.332$. The color stands for the ground state's angular momentum. The black region signifies no thermodynamically stable exciton states. The region below the constant $E_g/\epsilon'$ contours (dotted black lines) signify regions of phase diagram where no thermodynamically stable exciton is allowed at this value of bandgap. (b) Real part of ground state wavefunction in the momentum space in angular momentum channels $l = 0, -1, -2, -3$, respectively, for $V_0/\epsilon'=10, b k_0^2 = 1.8$. The wavefunctions are normalized such that the maximum value of the real part is 1. We find maximum probability density occurring near the minima of the Mexican hat bandstructure (at  $k \sim k_0$).}
    \label{fig:phase-diagram-single-exciton}
\end{figure}

One crucial difference between the chiral excitonic condensate presented here and other works focusing mainly on superconductivity (particle-particle pairing) \cite{Geier_chiral_2025, Chou_intravalley_2025, Joy_chiral_2025, Karuzin_bound_2026} is that our starting point is not a valley and spin polarized state, which would enforce a chiral state in the superconducting pairing channel due to Fermi statistics. Instead, for excitons, Berry curvature alone can lower the energy of a chiral state compared to its $s$-wave counterpart.

\paragraph*{Theory ---}
We begin with a band-insulating state with dispersions $\varepsilon_{c}(\mathbf{k})$ and $\varepsilon_{v}(\mathbf{k})$ for the conduction and valence bands, respectively, such that $|\varepsilon_{c}-\varepsilon_{v}| \geq E_g>0$. In the presence of long-range interaction $V(\mathbf{r}-\mathbf{r}') = \frac{1}{A} \sum_{\mathbf{q}}V(\mathbf{q})e^{i\mathbf{q}\cdot \mathbf{r}}$.

We define the ground state as a fully filled valence band $|\textrm{val}\rangle$. We additionally label $c_{\mathbf{k}}$ as conduction band annihilation operator and $d_{\mathbf{k}} = c^{\dagger}_{v,-\vec k}$ as hole annihilation (valence band electron creation) operator. The Hamiltonian, upto overall constants, is (see Appendix A for a complete derivation),
\begin{align}
   \notag H = &\sum_{\mathbf{k}}(\varepsilon_{c}(\mathbf{k}) -\mu) c^\dagger_{\mathbf{k}}c_{\mathbf{k}}-\sum_{\mathbf{k}}(\varepsilon_{v}(\mathbf{k})-\mu) d^\dagger_{-\mathbf{k}}d_{-\mathbf{k}} \\ &-\frac{1}{A}\sum_{\mathbf{k},\mathbf{k}',\mathbf{q}} V(\mathbf{q}) \Lambda_{cc}^{\mathbf{k}+\bf q, \mathbf{k}}\Lambda_{vv}^{\mathbf{k}'-\bf q, \mathbf{k}'}c^\dagger_{\mathbf{k}+\bf q} d^\dagger_{\mathbf{k}'-\bf q} d_{\bf k'} c_{\bf k}.
\end{align}
Here, $\Lambda_{ab}^{\bf k, \bf k'}= \langle u_a(\mathbf{k})|u_b(\mathbf{k}')\rangle $ is the form factor between the appropriate Bloch states, and we dropped an overall constant relating to the energy of the fully filled band. {As the conduction band is fully empty and the valence band is fully filled, for weak coupling, competing instabilities such as to charge or spin density waves are suppressed}.
Note the explicit negative sign indicating that the potential is in the direct binding channel of the Coulomb interaction (see Appendix A for derivation). In 2D, the Coulomb interaction is $V(\vec r_{2D}) = \frac{e^2}{4\pi \epsilon_0 \epsilon_r \sqrt{r_{2D}^2 + d_{\rm layer}^2}}$. After Fourier transforming and taking the effects of gates into account, $V(q) = \frac{2\pi e^2}{\epsilon_0 \epsilon_r q} \tanh(d_{\rm gate}q) e^{-{d_{\rm layer} q}}$. For the relevant values of $q$, the quantity $d_{\rm layer} q \ll 1$, and we discard the exponential factor \cite{ho2009ground}. 

\paragraph*{Single exciton approximation ---}
We now study the properties of individual excitons. We construct the single direct exciton ansatz as there is approximately conserved particle-hole symmetry in the systems of interest, $|1X\rangle = \sum_{\bf k}A_{\bf k} c^\dagger_{\bf k}d^\dagger_{-\mathbf{k}}|\textrm{val}\rangle$ (here $A_\mathbf{k}$ is the exciton envelope function). We arrive at an approximate form of the Bethe-Salpeter equation \cite{rohlfing2000electron} by applying the usual Wick contractions (see Appendix A):
\begin{equation}\label{eq:bethe-salpeter-single}
    \begin{aligned}
    &(\varepsilon_c(\mathbf k) - \varepsilon_v(\mathbf k)) A_{\mathbf k} ~ -\\ &\frac{1}{A}\sum_{\mathbf k'} \left[V(\mathbf k - \mathbf k') \Lambda_{cc}^{\mathbf k,\mathbf k'} \Lambda_{vv}^{\mathbf k' \mathbf k} A_{\mathbf k'}\right] = E A_{\mathbf k}
    \end{aligned}
\end{equation}
When the interaction is strong, it can make the overall energy eigenvalue $E$ negative, giving rise to a thermodynamically stable bound state of an electron and a hole.

We find that the form factors induce chirality in the ground state exciton, i.e., $A_{\vec k} \sim A_l(k) e^{i l \theta}$ with a non-zero $l$. This can be understood from the fact that for a small change $\vec q$ in momentum, $\Lambda_{cc}^{\mathbf{k}, \mathbf{k}+\vec q} \approx 1 - i \vec A_{cc} (\vec k)\cdot \vec q$, which is identical to the small $\vec q$ expansion of $e^{-i \vec A(\vec k) \cdot \vec q}$, where $\vec A_{cc}(\vec k) = i \bra{u_c(\vec k)}\partial_{\vec k} \ket{u_c(\vec k)}$ is the Berry connection. The valence band form factor produces a phase of the same sign. In other words, the form factors introduce an effective Aharonov-Bohm phase as the interaction scatters the electron-hole pair in the momentum space. 
Without this phase, we show (in the Appendix) that the $s$-wave exciton is always the ground state for the equivalent problem of a 2D hydrogen atom in a magnetic field.

Since the exciton is a boson, we treat condensation by examining the gap equation \cite{Jerome_Excitonic_1967},
\begin{equation}\label{eq:gap-eqn}
    \Delta_{\vec k} = \frac{1}{A} \sum_{\vec q}
    \mathcal{V}_{\mathbf{k},\mathbf{q}}
    \frac{\left[1 - f(E_1(\vec k+\vec q)) -f(E_2 (\vec k+\vec q)) \right]}{(E_1(\vec k+\vec q) + E_2({\vec k+\vec q}))} \Delta_{\vec k + \vec q}
\end{equation}
with $\mathcal{V}_{\mathbf{k},\mathbf{q}} = \left[V(\vec q)\Lambda_{cc}^{\vec k + \vec q, \vec k} \Lambda_{vv}^{\vec k, \vec k+\vec q}\right]$. Here the excitation energy of the Bogoliubov-like quasiparticles is given by
\begin{align}
   E_{1,2}(\vec k) = &\sqrt{|\Delta_{\vec k}|^2 + \varepsilon_{cv}^2/4} \pm ((\varepsilon_c (\vec k) +\varepsilon_v(\vec k))/2 - \mu),
\end{align}
where $\varepsilon_{cv} = \varepsilon_c(\mathbf{k})-\varepsilon_{v}(\mathbf{k})$. {  In appendix C, we demonstrate the symmetry of the single excitonic ground state tracks the symmetry of the condensate. Namely, if the Bethe-Salpeter equation (Eq.~\eqref{eq:bethe-salpeter-single}) describing a single exciton supports a bound state in some angular momentum channel, the channel corresponding to the ground state will host the leading instability in the linearized version of the gap equation (see Eq.~\eqref{eq:linearized-gap-eqn}).}

It is instructive to contrast the excitonic gap equation with that for superconductivity \cite{Bardeen1957,Chen_Review_2024}. Unlike the Cooper problem, a sign-changing interaction is not required for a non-zero angular momentum solution \cite{Anderson_morel,Anderson_Morel_2,Balian1963}. Moreover, unlike superconductors where an infinitesimal attractive interaction is sufficient for pairing, here the repulsive interaction needs to be strong enough to overcome the bandgap (i.e., a threshold) in order to produce a thermodynamically stable excitonic insulator.

Below, we study the ground state properties in a toy model with a double-well (``Sombrero") dispersion (see Fig.~\ref{fig:mexican_hat_cartoon}(a)), followed by a more realistic model of rhombohedral tetralayer graphene.

\paragraph*{Toy model ---}
Here we consider a toy model with a Mexican-hat like dispersion with tunable bandgap, inspired by the low-energy dispersion of a single valley in bernal bilayer graphene and a wide range of TMD materials \cite{Demirci_structural_2017},
\begin{equation}
    \varepsilon_c(\vec k) = -\varepsilon_v(\vec k) = \epsilon' \frac{(k^2-k_0^2)^2}{2 k_0^4} + \frac{E_g}{2},
\end{equation}
and set $\mu = 0$. Here, the minima of the Mexican hat at $k_0$, the rigidity $\epsilon'$ and the bandgap $E_g$ can be independently tuned.

\begin{figure}[t]
    \centering
    \includegraphics[width=0.99\linewidth]{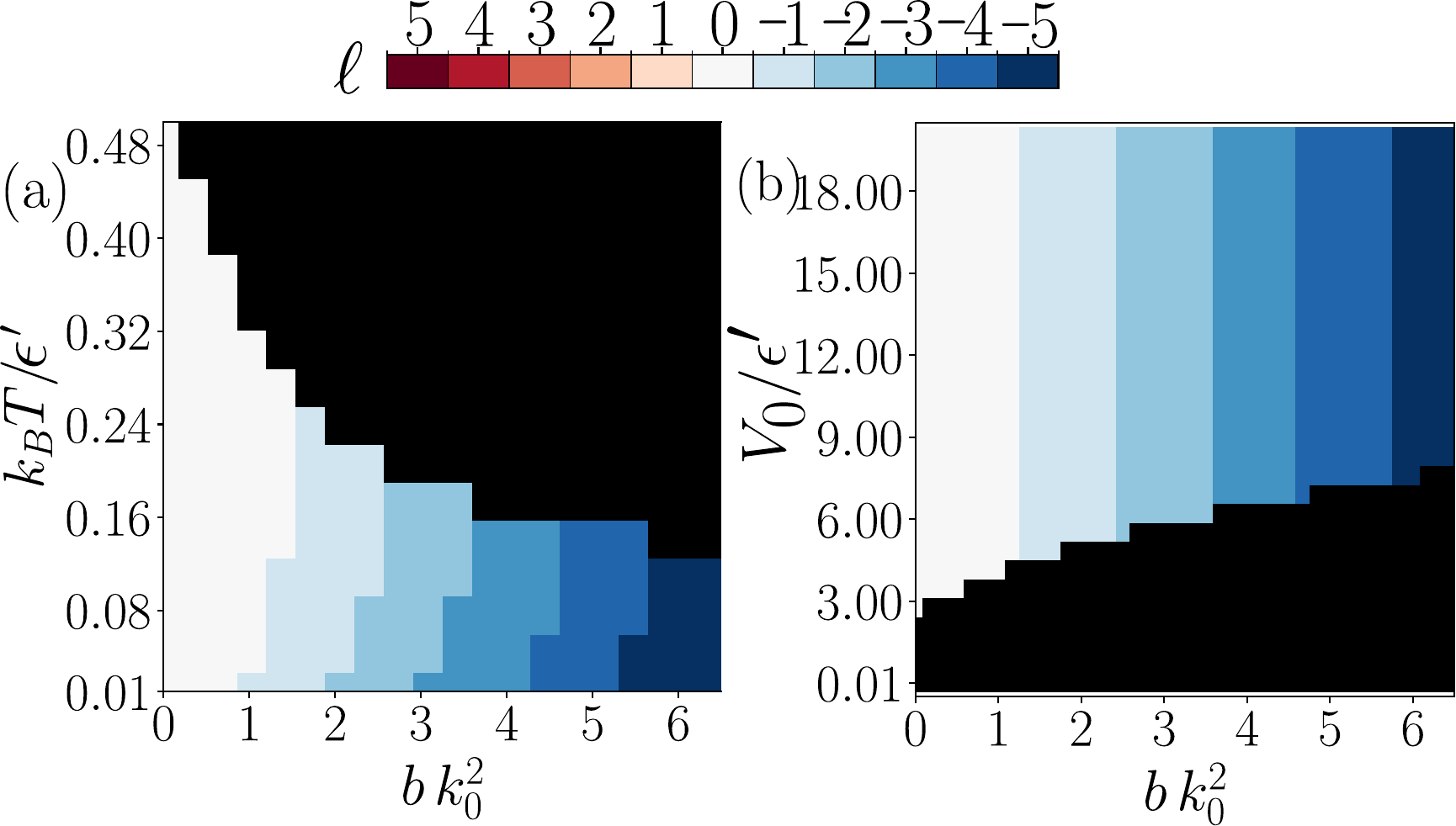}
    \caption{Phase diagram of the excitonic condensate. (a) The leading instability of the linearized gap equation as a function of temperature and Berry flux ($bk_0^2$), at a fixed interaction strength $V_0/\epsilon'= 10$. The black region on top stands for the region where the largest eigenvalue is less than 1, signifying lack of excitonic condensate.  (b) The angular momentum channel where the dominant instability occurs, computed by finding the leading eigenvalue of the kernel in the linearized gap equation. We have set temperature $k_B T/\epsilon' = 0.1 $. In both the plots, we have set $d = 1000/k_0$ and bandgap $E_g/\epsilon' = 0.1$.}
    \label{fig:condensate-phase-diagram}
\end{figure}

For simplicity, we consider a band with broken time-reversal symmetry, that hosts a uniform distribution of Berry curvature $b$, giving rise to Landau-level-like form factors \cite{Tan_parent_2024},
\begin{equation}
\begin{aligned}
    \Lambda_{cc}^{\vec k, \vec k'} &= e^{-\frac{1}{4}( |b| (\vec k - \vec k')^2 + 2i b (\vec k \times \vec k'))}\\
    \Lambda_{vv}^{\vec k, \vec k'} &= e^{-\frac{1}{4}( |b| (\vec k - \vec k')^2 - 2i b (\vec k \times \vec k'))}.
\end{aligned}
\end{equation}

The Berry flux enclosed by the annulus of the Mexican hat is, $\Phi = \pi k_0^2 b$.
We consider a gate-screened repulsive Coulomb interaction between electrons,
$V(k) = \frac{V_0 \tanh{(k d_{\rm gate})}}{k_0 k} $.

As the Hamiltonian has a continuous rotational symmetry, we can write solutions in individual angular momentum channels, $A_{\vec k} = \sum_{l} \psi_l(k) e^{i l \theta}$ (see Fig.~\ref{fig:phase-diagram-single-exciton}(b)), decomposing the BSE along those channels; recasting it in terms of dimensionless momentum $\kappa = k/k_0$, gap $E_g/\epsilon'$ and interaction $V_0/\epsilon'$,
\begin{equation}\label{eq:toy-mexican-non-dim-equation}
   \begin{aligned}
    \left[\frac{E_g}{\epsilon'} + (\kappa^2-1)^2 \right]\psi_l(\kappa) - \frac{V_0}{\epsilon'}  \frac{\Delta \kappa}{2\pi} \sum_{\kappa_2=0}^\infty  \kappa_2 &\tilde{V}_l(\kappa,\kappa_2)\psi_l (\kappa_2)\\  &= \frac{E}{\epsilon'} \psi_l(\kappa),
\end{aligned}
\end{equation}
where $\Delta \kappa$ is the effective area
(grid spacing), and $\tilde{V}_l(\kappa,\kappa_2)$ is the dimensionless interaction in the $l$-th angular momentum channel,

\begin{equation}
\begin{aligned}
    \tilde{V_l}(\kappa_1,\kappa_2) = \int_{0}^{2\pi}\frac{d\chi}{2\pi} &{e^{i l \chi} e^{-\frac{b k_0^2}{2} \left[{|\boldsymbol \kappa_1 - \boldsymbol \kappa_2|^2} + 2 i \kappa \kappa_2 \sin\chi\right]}}\\
    &\quad \times \frac{\tanh{\left(k_0 d_{\rm gate} |\boldsymbol \kappa_1 - \boldsymbol \kappa_2|\right)}}{|\boldsymbol \kappa_1 - \boldsymbol \kappa_2|}.
\end{aligned}    
\end{equation}

with $|\boldsymbol \kappa_1 - \boldsymbol \kappa_2| = \sqrt{\kappa_1^2 + \kappa_2^2 - 2\kappa_1 \kappa_2 \cos \chi}$.

We find the eigenvalues at a fixed bandgap, and obtain the phase diagram by comparing the ground state energies in each angular momentum channel. We find that at the limit of small interactions and small bandgap, the system undergoes phase transition to angular momentum $l$ when the Berry flux enclosed by the annulus at $k_0$ crosses integer multiples of $\pi$, i.e., $\Phi = \Phi_n \approx n \pi$ (Fig.~\ref{fig:phase-diagram-single-exciton}(a)).

Moreover, to study the leading instability in the excitonic condensate, we consider the linearized gap equation,
\begin{equation}\label{eq:linearized-gap-eqn}
    \Delta_{\vec k} = \frac{1}{A} \sum_{\vec q}  \mathcal{V}_{\mathbf{k},\mathbf{q}} \frac{\Delta_{\vec k + \vec q} \tanh{\left(\frac{\varepsilon_c(\vec k +\vec q)-\varepsilon_v(\vec k+\vec q)}{4 k_B T}\right)}}{\varepsilon_c(\vec k+\vec q)-\varepsilon_v(\vec k+\vec q)} ,
\end{equation}
and compare the eigenvalues of the kernel of this equation at a finite temperature to find the leading excitonic instability. We find that the angular momentum of the leading excitonic instability closely tracks the single excitonic ground state described by the BSE, as we show in Fig.~\ref{fig:condensate-phase-diagram}(a), (b). We also find that the order of magnitude of $T_c$ varies between $0.2 \epsilon'/k_B$ to $0.5 \epsilon'/k_B$.

In this model we can independently tune the interaction, the curvature of the band (set by $\epsilon'$) and the band gap.
Now, we study a more realistic model of tetralayer graphene, where the interaction strength is set by fundamental constants of nature, and the band curvature and the band gap are simultaneously tuned by a displacement field. Moreover, due to trigonal warping, rotational symmetry is broken, and different angular momentum channels mix to form the energy eigenstates.

\begin{figure}[!t]
    \centering
    \includegraphics[width=0.98\linewidth]{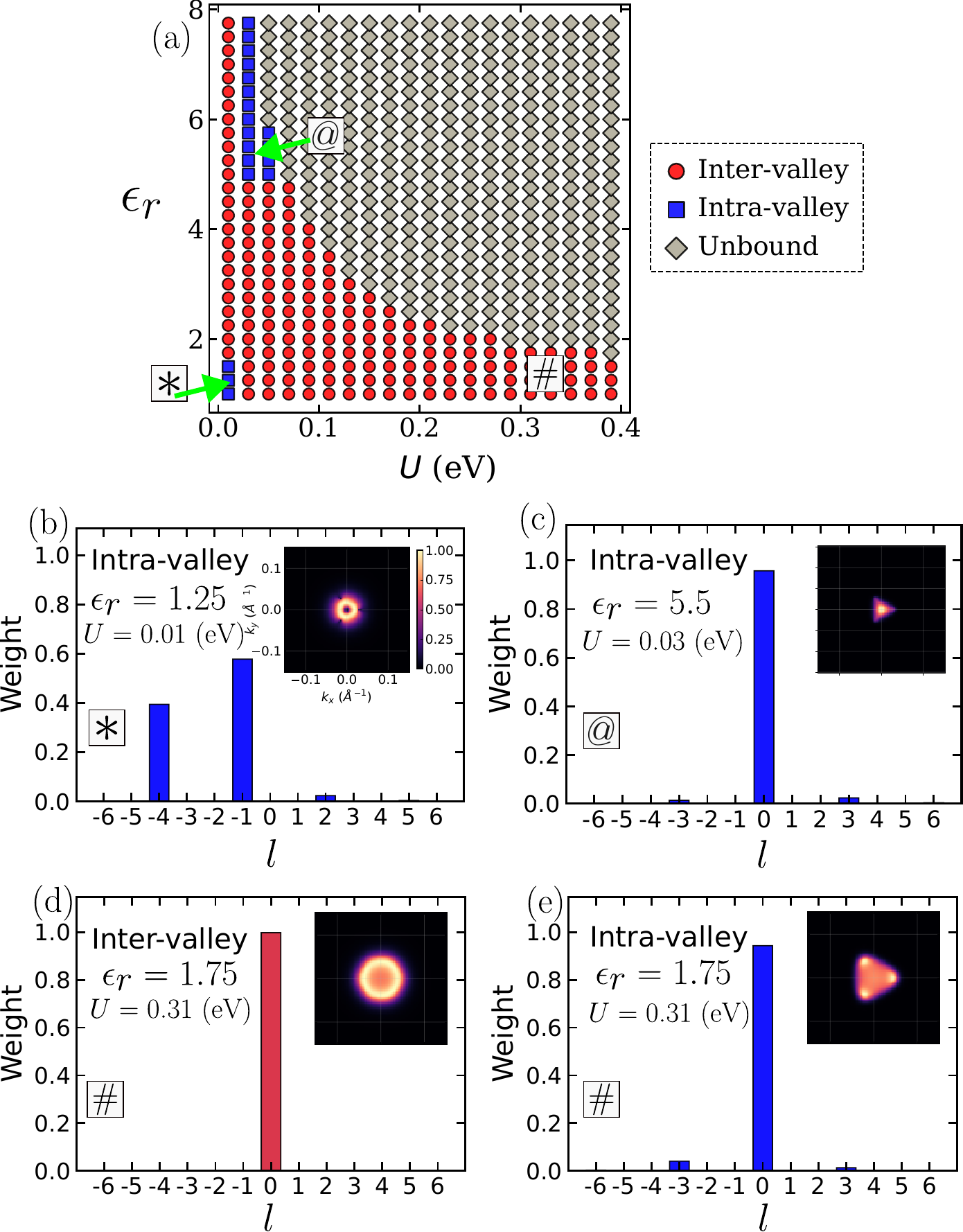}
    \caption{(a) Phase diagram of a single exciton in tetralayer graphene as a function of  field $U$ and dielectric constant $\epsilon_r$ at a gate distance $d_{\rm gate} = 50$nm. The blue and the red colors stand for intra-valley and inter-valley ground states. The black points at large band gap and large dielectric constant stand for thermodynamically unstable states. (b) Shows the angular momentum decomposition in the predominantly chiral $p$-wave intra-valley ground state for $\epsilon_r = 1.25, U = 0.01 eV$, and the inset shows the relative probability density $|A_{\vec k}|^2$. In panel (c) we display the same quantity for $\epsilon_r = 1.75, U = 0.03$ (eV) where the ground state is intra-valley, and has overall chirality, but $s$-wave is the dominant component. In (d), we plot the angular momentum decomposition of the achiral inter-valley exciton, which is the ground state for $\epsilon_r = 1.75, U = 0.31$ (eV), and in panel (e) we plot the wavefunction and angular momentum decomposition for the thermodynamically unstable intra-valley exciton with $\epsilon_r = 1.75, U = 0.31$ (eV).}
    \label{fig:graphene_tetralayer}
\end{figure}

\paragraph*{Rhombohedral tetralayer graphene---} 
We consider a 8-band matrix Hamiltonian for $K$ valley of rhombohedral tetralayer graphene as introduced in Ref.~\cite{MacDonald_ABCtrilayer_2010} (details in Appendix D), with the gate-screened Coulomb interaction between electrons and holes, and solve Eq.~\eqref{eq:bethe-salpeter-single}. The solutions for $K'$ valley follow from TRS. Below, we will argue that excitons can spontaneously undergo time-reversal symmetry breaking via a Stoner-like mechanism, and thus we focus on the case where the electron is in $K$ valley, and the hole can be in either $K$ valley (intravalley exciton) or $K'$ valley (intervalley exciton). In the intravalley exciton, the complex phases of $\Lambda_{cc}^{\vec k + \vec q, \vec k}$ and $ \Lambda_{vv}^{\vec k, \vec k+\vec q}$ add up, favoring a chiral ground state. In comparison, the phases cancel out in the intervalley exciton, resulting in a predominantly $s$-wave state with equal population of finite angular momentum contribution of either sign.

The phase diagram for a fixed value of gate distance is presented in Fig.~\ref{fig:graphene_tetralayer}(a). Here, the control knobs are the displacement field, which sets the band gap, and the dielectric screening $\epsilon_r$, which can be controlled by the encapsulating environment \cite{laturia2018dielectric}. If the dielectric constant is too large, the interaction is heavily screened, and the system cannot form excitonic bound state even at a reasonably small displacement field. 
As the trigonal warping breaks rotational invariance, an excitonic state does not have a definite angular momentum, unlike the toy models studied previously. However, we can still find the dominant angular momentum channel for a particular ground state, as we show in Figs.~\ref{fig:graphene_tetralayer}(c)-(e). For small values of dielectric constant and displacement field, the excitonic envelope function shows (whose absolute magnitude is depicted in the inset of Fig.~\ref{fig:graphene_tetralayer}(c)) signs of several sign changes, indicating $p$-wave like structure. By projecting onto angular momenta (see Appendix D), this is revealed by a dominant $l=-1$ contribution (Fig.~\ref{fig:graphene_tetralayer}(c)), followed by the contribution of the $l=-4$ channel, as trigonal warping couples $l$ with $l\pm3$ for a lattice with three-fold rotational symmetry.
We find that for very small displacement fields and dielectric screening ($\epsilon_r < 2$), the ground state is intra-valley, and is predominantly chiral. As either of dielectric constant or the displacement field are increased, eventually the achiral intervalley exciton becomes the ground state. While intravalley excitons are again observed for large $\epsilon_r$, they are predominantly $s$-wave, with little overall chirality due to imbalance between $l=\pm 3$ components.

At large displacement fields, the Coulomb energy cannot anymore overcome the gap between the valence band and the conduction band, and while in-gap chiral excitons exist, they are no longer thermodynamically stable, and require optical pumping to form an excitonic condensate.

\paragraph*{Discussion---}

In this work we showed through a combination of toy models and direct calculations on rhombohedral tetralayer graphene the effects of Berry phases on the problem of exciton condensation. The quantum geometry of Bloch bands was introduced using complex form factors, which in turn favored bound states at finite angular momentum. The reason for this relies on the structure of the form factors, $\Lambda_{\mathbf{k},\mathbf{q}}\approx 1-i\mathbf{A}_{\mathbf{k}}\cdot \mathbf{q}$, which add an odd in $\mathbf{q}$ dressing to electron-electron interactions for topological bands. In the limit of small bandgap and weak interaction strength, a cascade of phase transitions from angular momentum $l$ to $l+1$ occurs when the Berry flux within the rim of the Mexican hat crosses the value $(l+1) \pi$, an effect that is also observed in the leading instability obtained from the linearized gap equation describing the chiral excitonic phase (Fig.~\ref{fig:condensate-phase-diagram}(a)). The Berry curvature alone can switch the ground state from $s$-wave to chiral $p$-wave, an effect not observed in chiral superconductors, where the non-$s$-wave nature is enforced by Fermi statistics (if the underlying Fermi liquid is already time-reversal symmetry broken).
At stronger interactions, we find a first order transition between the $s$-wave and $d$-wave exciton. {Further, we apply this mechanism to a model of rhombohedral tetralayer graphene, and demonstrate that it too hosts chiral excitonic bound states, based on these simple considerations.}
We now address whether the condensate may favor spontaneous valley polarization, even if the underlying normal state is time-reversal symmetric. 
{ 
An analysis similar to that in Ref.~\cite{Wu_Macdonald2015} allows us to construct the free-energy functional in terms of the exciton density (see Appendix G),
\begin{equation}
\begin{aligned}
F &= -a (n_1 + n_2) + \left(g_H + \frac{g_X}{2}\right) (n_1 + n_2)^2 + \\ &\frac{g_X}{2}(n_1 - n_2)^2
= - a N + \left(g_H + \frac{g_X}{2}\right) N^2 + \frac{g_X}{2} \Delta^2.
\end{aligned}
\end{equation}
Here, $n_{1,2}$ are the densities of the excitons in the two valleys. $N = n_1 + n_2$ is the total excitonic density and $\Delta = (n_1 - n_2)$ denotes valley-polarization of excitons. Here, $g_X$ is an exchange term generated by boson-boson interactions, and $g_X < 0$ favors valley polarization. In appendix G, we argue that under small density of excitons, this term can indeed be negative, favoring valley polarization of the condensate. Under small densities, the condition for valley-polarization has no threshold as the free energy is only quadratic in $\Delta$.
At large excitonic densities there will be $\mathcal{O}(\Delta^4)$ terms which will set the upper limit of valley polarization.
}

Finally, let us discuss experimental signatures of the thermodynamically stable chiral excitons.  We expect the excitonic condensate to display a valley thermal Hall effect and a Hall-like response in Coulomb drag \cite{Moon_exciton_2025}. If electrons are injected into this fluid, a Magnus effect may arise as a result of chirality \cite{Ikegami_chiral_2013}.
As the orbital angular momentum of the chiral excitons couple strongly with an out-of-plane magnetic field, it can break the valley degeneracy of the chiral excitons, leading to hysteresis of thermal Hall response under a small magnetic field, an effect that should be absent in $s$-wave excitons. In a magnetic field, the energy balance between intra- and inter-valley states will shift, due to a narrowing of the inter-valley gap~\cite{Dinh_Dery2025,PhysRevB.102.195403,PhysRevB.99.115439}. However, due to the coupling of the magnetic field to the finite-$l$ magnetic moment of the exciton, a region in the phase is expected where the magnetic field \textit{enhances} intravalley pairing. This will also be reflected as an increase in the condensate $T_c$.
{  The broken time-reversal symmetry and chirality of the excitonic condensates can likewise be probed using STM, through impurity bound states and quasiparticle interference~\cite{Kwon_defect-bound_2026}, analogous to the recent proposal in unconventional superconductors~\cite{Panigrahi_tomographic_2026}, and even normal transition-metal dichalcogenides (TMDs) \cite{holbrook2026real}.}

{ Before closing, we remark that the inter-valley exciton involves $\langle c^{\dagger}_{\vec K+\vec k_1, c} c_{\vec K' + \vec k_2, v}\rangle$, which is simultaneously a intervalley-coherent state, and also a charge density wave. Here, we examined the competition between this state and the chiral excitonic codensate (which does not break translation symmetry), and showed this in Fig.~\ref{fig:graphene_tetralayer}(a).

In this work, we neglected the effects of fluctuations beyond mean-field.}
{  It was recently demonstrated~\cite{Liu_optical_2025} that experimental values of $T_c$ in van der Waals materials can be as high as $75$K, which is about $30\%$ of the mean-field estimate, indicating the qualitative validity of our results.}  It is also natural to ask whether increasing the number of layers in rhombohedral graphene will further stabilize the chiral excitonic states, given the important role of Berry curvature, as evinced here. {This also serves to contrast our case with  TMD platforms, as neither the gap nor the Berry flux, are easily tunable in the latter. We also note that our conclusions will be particularly relevant for moir\'e systems, where the narrow dispersion and strong dependence on stacking order may stabilize new excitonic states \cite{huang2022excitons,huang2023mott}. It should be possible to polarize, via excitons, mini-valley as well as parent valley states with the additional degrees of freedom in moir\'e superlattices.}
This idea, and the full phase diagram involving the effects of fluctuations, as well as the details of the valley-symmetry breaking in realistic systems, will be addressed in future work. It has been theoretically suggested that exciton-like condensates generated via spontaneous breaking of subsystem symmetries can lead to anisotropic marginal Fermi liquid behavior in a system with a large number of layers \cite{Panigrahi_non-fermi_2025}. It would be intriguing to study the analogous effect for chiral excitons.

\paragraph*{Acknowledgements ---}
We are grateful to Andrey Chubukov, Daniele Guerci, Long Ju, Elio K\"onig, Ryan Lee, Leonid Levitov, Xiaohui Liu, Siddharth Parameswaran, Gurkirat Singh, Pavel Volkov, and Bo Zou for insightful discussions. This work was initiated at the Aspen Center for Physics, which is supported by National Science Foundation grant PHY-2210452. DK was supported by the Abrahams Postdoctoral Fellowship of the Center for Materials Theory, Rutgers University.

\bibliography{references.bib}

@article{Yang_analytic_1991,
 author = {Yang, X. L. and Guo, S. H. and Chan, F. T. and Wong, K. W. and Ching, W. Y.},
 issn = {1094-1622},
 journal = {Phys. Rev. A},
 month = {February},
 number = {3},
 pages = {1186--1196},
 publisher = {American Physical Society (APS)},
 title = {{Analytic} solution of a two-dimensional hydrogen atom. {I}. {Nonrelativistic} theory},
 url = {http://dx.doi.org/10.1103/PhysRevA.43.1186},
 volume = {43},
 year = {1991}
}

@article{Nandi_drag_2012,
 author = {Nandi, D. and Finck, A. D. K. and Eisenstein, J. P. and Pfeiffer, L. N. and West, K. W.},
 issn = {1476-4687},
 journal = {Nature},
 month = {August},
 number = {7412},
 pages = {481--484},
 publisher = {Springer Science and Business Media LLC},
 title = {{Exciton} condensation and perfect {Coulomb} drag},
 url = {http://dx.doi.org/10.1038/nature11302},
 volume = {488},
 year = {2012}
}

@article{qi2023thermodynamic,
 author = {Qi, Ruishi and Joe, Andrew Y. and Zhang, Zuocheng and Zeng, Yongxin and Zheng, Tiancheng and Feng, Qixin and Xie, Jingxu and Regan, Emma and Lu, Zheyu and Taniguchi, Takashi and Watanabe, Kenji and Tongay, Sefaattin and Crommie, Michael F. and MacDonald, Allan H. and Wang, Feng},
 issn = {2041-1723},
 journal = {Nat. Commun.},
 month = {December},
 number = {1},
 publisher = {Springer Science and Business Media LLC},
 title = {{Thermodynamic} behavior of correlated electron-hole fluids in van der {Waals} heterostructures},
 url = {http://dx.doi.org/10.1038/s41467-023-43799-7},
 volume = {14},
 year = {2023}
}

@article{Jerome_Excitonic_1967,
  title = {Excitonic Insulator},
  author = {J\'erome, D. and Rice, T. M. and Kohn, W.},
  journal = {Phys. Rev.},
  volume = {158},
  issue = {2},
  pages = {462--475},
  numpages = {0},
  year = {1967},
  month = {Jun},
  publisher = {American Physical Society},
  doi = {10.1103/PhysRev.158.462},
  url = {https://link.aps.org/doi/10.1103/PhysRev.158.462}
}

@article{Tan_parent_2024,
 author = {Tan, Tixuan and Devakul, Trithep},
 issn = {2160-3308},
 journal = {Phys. Rev. X},
 month = {November},
 number = {4},
 pages = {041040},
 publisher = {American Physical Society (APS)},
 title = {{Parent} {Berry} {Curvature} and the {Ideal} {Anomalous} {Hall} {Crystal}},
 url = {http://dx.doi.org/10.1103/PhysRevX.14.041040},
 volume = {14},
 year = {2024}
}

@article{Demirci_structural_2017,
 author = {Demirci, S. and Avazl{\i}, N. and Durgun, E. and Cahangirov, S.},
 issn = {2469-9969},
 journal = {Phys. Rev. B},
 month = {March},
 number = {11},
 pages = {115409},
 publisher = {American Physical Society (APS)},
 title = {{Structural} and electronic properties of monolayer group {III} monochalcogenides},
 url = {http://dx.doi.org/10.1103/PhysRevB.95.115409},
 volume = {95},
 year = {2017}
}

@article{MacDonald_ABCtrilayer_2010,
 author = {Zhang, Fan and Sahu, Bhagawan and Min, Hongki and MacDonald, A. H.},
 issn = {1550-235X},
 journal = {Phys. Rev. B},
 month = {July},
 number = {3},
 pages = {035409},
 publisher = {American Physical Society (APS)},
 title = {{Band} structure of ${ABC}$ -stacked graphene trilayers},
 url = {http://dx.doi.org/10.1103/PhysRevB.82.035409},
 volume = {82},
 year = {2010}
}

@article{Nguyen_quantum_2025,
 author = {Nguyen, Phuong X. and Chaturvedi, Raghav and Zou, Bo and Watanabe, Kenji and Taniguchi, Takashi and MacDonald, Allan H. and Mak, Kin Fai and Shan, Jie},
 issn = {1476-4660},
 journal = {Nat. Mater.},
 month = {August},
 number = {1},
 pages = {42--48},
 publisher = {Springer Science and Business Media LLC},
 title = {{Quantum} oscillations in a dipolar excitonic insulator},
 url = {http://dx.doi.org/10.1038/s41563-025-02334-3},
 volume = {25},
 year = {2025}
}

@article{Panigrahi_non-fermi_2025,
  title = {{Non}-{Fermi} {Liquids} from {Subsystem} {Symmetry} {Breaking} in {van} der {Waals} {Multilayers}},
  author = {Panigrahi, Archisman and Kumar, Ajesh},
  journal = {Phys. Rev. Lett.},
  volume = {134},
  issue = {23},
  pages = {236502},
  numpages = {7},
  year = {2025},
  month = {Jun},
  publisher = {American Physical Society},
  doi = {10.1103/v6r7-4ph9},
  url = {https://link.aps.org/doi/10.1103/v6r7-4ph9}
}

@misc{Karuzin_bound_2026,
      title={Bound States from {Berry} {Curvature} and {Chiral} {Superconductivity}}, 
      author={Daniil Karuzin and Zhiyu Dong and Leonid Levitov},
      year={2026},
      eprint={2601.08055},
      archivePrefix={arXiv},
      primaryClass={cond-mat.mes-hall},
      url={https://arxiv.org/abs/2601.08055}, 
}

@article{Geier_chiral_2025,
 author = {Geier, Max and Davydova, Margarita and Fu, Liang},
 issn = {2041-1723},
 journal = {Nat. Commun.},
 month = {December},
 number = {1},
 publisher = {Springer Science and Business Media LLC},
 title = {{Chiral} and topological superconductivity in isospin polarized multilayer graphene},
 url = {http://dx.doi.org/10.1038/s41467-025-66902-6},
 volume = {17},
 year = {2025}
}

@article{Chou_intravalley_2025,
 author = {Chou, Yang-Zhi and Zhu, Jihang and Das Sarma, Sankar},
 issn = {2469-9969},
 journal = {Phys. Rev. B},
 month = {May},
 number = {17},
 pages = {174523},
 publisher = {American Physical Society (APS)},
 title = {{Intravalley} spin-polarized superconductivity in rhombohedral tetralayer graphene},
 url = {http://dx.doi.org/10.1103/PhysRevB.111.174523},
 volume = {111},
 year = {2025}
}

@article{Sundaram_wave-packet_1999,
 author = {Sundaram, Ganesh and Niu, Qian},
 issn = {1095-3795},
 journal = {Phys. Rev. B},
 month = {June},
 number = {23},
 pages = {14915--14925},
 publisher = {American Physical Society (APS)},
 title = {{Wave-packet} dynamics in slowly perturbed crystals: {Gradient} corrections and {Berry-phase} effects},
 url = {http://dx.doi.org/10.1103/physrevb.59.14915},
 volume = {59},
 year = {1999}
}

@article{Joy_chiral_2025,
 author = {Joy, Sandeep and Levitov, Leonid and Skinner, Brian},
 issn = {1079-7114},
 journal = {Phys. Rev. Lett.},
 month = {December},
 number = {25},
 pages = {256502},
 publisher = {American Physical Society (APS)},
 title = {{Chiral} {Wigner} {Crystal} {Phases} {Induced} by {Berry} {Curvature}},
 url = {http://dx.doi.org/10.1103/h5hy-jh6m},
 volume = {135},
 year = {2025}
}

@article{Peotta_superfluidity_2015,
 author = {Peotta, Sebastiano and Törmä, Päivi},
 issn = {2041-1723},
 journal = {Nat. Commun.},
 month = {November},
 number = {1},
 publisher = {Springer Science and Business Media LLC},
 title = {{Superfluidity} in topologically nontrivial flat bands},
 url = {http://dx.doi.org/10.1038/ncomms9944},
 volume = {6},
 year = {2015}
}

@article{Chen_Review_2024,
  title = {When superconductivity crosses over: From BCS to BEC},
  author = {Chen, Qijin and Wang, Zhiqiang and Boyack, Rufus and Yang, Shuolong and Levin, K.},
  journal = {Rev. Mod. Phys.},
  volume = {96},
  issue = {2},
  pages = {025002},
  numpages = {54},
  year = {2024},
  month = {May},
  publisher = {American Physical Society},
  doi = {10.1103/RevModPhys.96.025002},
  url = {https://link.aps.org/doi/10.1103/RevModPhys.96.025002}
}

@article{yu2025quantum,
  title={Quantum geometry in quantum materials},
  author={Yu, Jiabin and Bernevig, B Andrei and Queiroz, Raquel and Rossi, Enrico and T{\"o}rm{\"a}, P{\"a}ivi and Yang, Bohm-Jung},
  journal={npj Quantum Materials},
  volume={10},
  number={1},
  pages={101},
  year={2025},
  publisher={Nature Publishing Group UK London}
}

@article{Wu_exciton_2015,
 author = {Wu, Fengcheng and Qu, Fanyao and MacDonald, A. H.},
 issn = {1550-235X},
 journal = {Phys. Rev. B},
 month = {February},
 number = {7},
 pages = {075310},
 publisher = {American Physical Society (APS)},
 title = {{Exciton} band structure of monolayer ${\mathrm{MoS}}_{2}$},
 url = {http://dx.doi.org/10.1103/PhysRevB.91.075310},
 volume = {91},
 year = {2015}
}

@article{Zhou_Berry_2015,
  author = {Zhou, Jianhui and Shan, Wen-Yu and Yao, Wang and Xiao, Di},
  title = {{Berry Phase Modification to the Energy Spectrum of Excitons}},
  journal = {Phys. Rev. Lett.},
  volume = {115},
  pages = {166803},
  year = {2015},
  doi = {10.1103/PhysRevLett.115.166803}
}

@article{Kwan_excitonic_2022,
 author = {Kwan, Yves H. and Hu, Yichen and Simon, Steven H. and Parameswaran, S. A.},
 issn = {2469-9969},
 journal = {Phys. Rev. B},
 month = {June},
 number = {23},
 pages = {235121},
 publisher = {American Physical Society (APS)},
 title = {{Excitonic} fractional quantum {Hall} hierarchy in moiré heterostructures},
 url = {http://dx.doi.org/10.1103/PhysRevB.105.235121},
 volume = {105},
 year = {2022}
}

@article{Xie_topological_2024,
 author = {Xie, Hong-Yi and Ghaemi, Pouyan and Mitrano, Matteo and Uchoa, Bruno},
 issn = {1091-6490},
 journal = {Proceedings of the National Academy of Sciences},
 month = {August},
 number = {35},
 publisher = {Proceedings of the National Academy of Sciences},
 title = {{Theory} of topological exciton insulators and condensates in flat {Chern} bands},
 url = {http://dx.doi.org/10.1073/pnas.2401644121},
 volume = {121},
 year = {2024}
}

@article{Keldysh_instability_1965,
  author = {Keldysh, L. V. and Kopaev, Yu. V.},
  title = {Possible Instability of the Semimetallic State toward Coulomb Interaction},
  journal = {Soviet Physics Solid State},
  volume = {6},
  pages = {2219--2224},
  year = {1965}
}

@article{Kohn_exciton_1967,
 author = {Kohn, W.},
 issn = {0031-9007},
 journal = {Phys. Rev. Lett.},
 month = {August},
 number = {8},
 pages = {439--442},
 publisher = {American Physical Society (APS)},
 title = {{Excitonic} {Phases}},
 url = {http://dx.doi.org/10.1103/PhysRevLett.19.439},
 volume = {19},
 year = {1967}
}

@article{Halperin_Rice_1968,
  author = {Halperin, B. I. and Rice, T. M.},
  title = {Possible Anomalies at a Semimetal-Semiconductor Transition},
  journal = {Reviews of Modern Physics},
  volume = {40},
  pages = {755--766},
  year = {1968},
  doi = {10.1103/RevModPhys.40.755}
}

@article{Xiao-Chang-Niu_Berry,
  title = {{Berry phase effects on electronic properties}},
  author = {Xiao, Di and Chang, Ming-Che and Niu, Qian},
  journal = {Rev. Mod. Phys.},
  volume = {82},
  issue = {3},
  pages = {1959--2007},
  numpages = {0},
  year = {2010},
  month = {Jul},
  publisher = {American Physical Society},
  doi = {10.1103/RevModPhys.82.1959},
  url = {https://link.aps.org/doi/10.1103/RevModPhys.82.1959}
}

@misc{Vanderbilt_berry_2018,
  author = {Vanderbilt, David},
  title = {{Berry Phases in Electronic Structure Theory: Electric Polarization, Orbital Magnetization and Topological Insulators}},
  publisher = {Cambridge University Press},
  year = {2018}
}

@article{Hu_quantum_metric_2022,
 author = {Hu, Xiang and Hyart, Timo and Pikulin, Dmitry I. and Rossi, Enrico},
 issn = {2469-9969},
 journal = {Phys. Rev. B},
 month = {April},
 number = {14},
 pages = {L140506},
 publisher = {American Physical Society (APS)},
 title = {{Quantum-metric-enabled} exciton condensate in double twisted bilayer graphene},
 url = {http://dx.doi.org/10.1103/PhysRevB.105.L140506},
 volume = {105},
 year = {2022}
}

@article{Yao_exciton_2008,
 author = {Yao, Wang and Niu, Qian},
 issn = {1079-7114},
 journal = {Phys. Rev. Lett.},
 month = {September},
 number = {10},
 pages = {106401},
 publisher = {American Physical Society (APS)},
 title = {{Berry} {Phase} {Effect} on the {Exciton} {Transport} and on the {Exciton} {Bose-Einstein} {Condensate}},
 url = {http://dx.doi.org/10.1103/PhysRevLett.101.106401},
 volume = {101},
 year = {2008}
}

@article{julku_geometric_2016,
  title = {Geometric Origin of Superfluidity in the Lieb-Lattice Flat Band},
  author = {Julku, Aleksi and Peotta, Sebastiano and Vanhala, Tuomas I. and Kim, Dong-Hee and T\"orm\"a, P\"aivi},
  journal = {Phys. Rev. Lett.},
  volume = {117},
  issue = {4},
  pages = {045303},
  numpages = {6},
  year = {2016},
  month = {Jul},
  publisher = {American Physical Society},
  doi = {10.1103/PhysRevLett.117.045303},
  url = {https://link.aps.org/doi/10.1103/PhysRevLett.117.045303}
}

@article{Liang_band_2017,
  title = {{Band geometry, Berry curvature, and superfluid weight}},
  author = {Liang, Long and Vanhala, Tuomas I. and Peotta, Sebastiano and Siro, Topi and Harju, Ari and T\"orm\"a, P\"aivi},
  journal = {Phys. Rev. B},
  volume = {95},
  issue = {2},
  pages = {024515},
  numpages = {16},
  year = {2017},
  month = {Jan},
  publisher = {American Physical Society},
  doi = {10.1103/PhysRevB.95.024515},
  url = {https://link.aps.org/doi/10.1103/PhysRevB.95.024515}
}

@article{Srivastava_signatures_2015,
 author = {Srivastava, Ajit and Imamo\u{g}lu, Ata\c{c}},
 issn = {1079-7114},
 journal = {Phys. Rev. Lett.},
 month = {October},
 number = {16},
 pages = {166802},
 publisher = {American Physical Society (APS)},
 title = {{Signatures} of {Bloch-Band} {Geometry} on {Excitons}: {Nonhydrogenic} {Spectra} in {Transition-Metal} {Dichalcogenides}},
 url = {http://dx.doi.org/10.1103/PhysRevLett.115.166802},
 volume = {115},
 year = {2015}
}

@article{sun2022evidence,
 author = {Sun, Bosong and Zhao, Wenjin and Palomaki, Tauno and Fei, Zaiyao and Runburg, Elliott and Malinowski, Paul and Huang, Xiong and Cenker, John and Cui, Yong-Tao and Chu, Jiun-Haw and Xu, Xiaodong and Ataei, S. Samaneh and Varsano, Daniele and Palummo, Maurizia and Molinari, Elisa and Rontani, Massimo and Cobden, David H.},
 issn = {1745-2481},
 journal = {Nat. Phys.},
 month = {December},
 number = {1},
 pages = {94--99},
 publisher = {Springer Science and Business Media LLC},
 title = {{Evidence} for equilibrium exciton condensation in monolayer {WTe$_2$}},
 url = {http://dx.doi.org/10.1038/s41567-021-01427-5},
 volume = {18},
 year = {2021}
}

@article{Zeng_topological_excitation,
  title = {Topological Excitonic Insulators in Electron Bilayers Modulated by Twisted Hexagonal Boron Nitride},
  author = {Zeng, Yongxin and MacDonald, Allan H. and Wei, Nemin},
  journal = {Phys. Rev. Lett.},
  volume = {136},
  issue = {10},
  pages = {106602},
  numpages = {8},
  year = {2026},
  month = {Mar},
  publisher = {American Physical Society},
  doi = {10.1103/3s2h-5x77},
  url = {https://link.aps.org/doi/10.1103/3s2h-5x77}
}

@article{ma2021strongly,
 author = {Ma, Liguo and Nguyen, Phuong X. and Wang, Zefang and Zeng, Yongxin and Watanabe, Kenji and Taniguchi, Takashi and MacDonald, Allan H. and Mak, Kin Fai and Shan, Jie},
 issn = {1476-4687},
 journal = {Nature},
 month = {October},
 number = {7882},
 pages = {585--589},
 publisher = {Springer Science and Business Media LLC},
 title = {{Strongly} correlated excitonic insulator in atomic double layers},
 url = {http://dx.doi.org/10.1038/s41586-021-03947-9},
 volume = {598},
 year = {2021}
}

@article{wang2019evidence,
 author = {Wang, Zefang and Rhodes, Daniel A. and Watanabe, Kenji and Taniguchi, Takashi and Hone, James C. and Shan, Jie and Mak, Kin Fai},
 issn = {1476-4687},
 journal = {Nature},
 month = {October},
 number = {7776},
 pages = {76--80},
 publisher = {Springer Science and Business Media LLC},
 title = {{Evidence} of high-temperature exciton condensation in two-dimensional atomic double layers},
 url = {http://dx.doi.org/10.1038/s41586-019-1591-7},
 volume = {574},
 year = {2019}
}

@article{hu2022quantum,
  title={Quantum-metric-enabled exciton condensate in double twisted bilayer graphene},
  author={Hu, Xiang and Hyart, Timo and Pikulin, Dmitry I and Rossi, Enrico},
  journal={Phys. Rev. B},
  volume={105},
  number={14},
  pages={L140506},
  year={2022},
  publisher={APS}
}

@article{Guerci2025,
  title = {From Fractionalization to Chiral Topological Superconductivity in a Flat Chern Band},
  author = {Guerci, Daniele and Abouelkomsan, Ahmed and Fu, Liang},
  journal = {Phys. Rev. Lett.},
  volume = {135},
  issue = {18},
  pages = {186601},
  numpages = {8},
  year = {2025},
  month = {Oct},
  publisher = {American Physical Society},
  doi = {10.1103/zm39-dstj},
  url = {https://link.aps.org/doi/10.1103/zm39-dstj}
}

@article{Martinez2025,
  title = {Band Renormalization, Quarter Metals, and Chiral Superconductivity in Rhombohedral Tetralayer Graphene},
  author = {Parra-Mart\'{\i}nez, Guillermo and Jimeno-Pozo, Alejandro and Phong, V\~o Tiến and Sainz-Cruz, H\'ector and Kaplan, Daniel and Emanuel, Peleg and Oreg, Yuval and Pantale\'on, Pierre A. and Silva-Guill\'en, Jos\'e \'Angel and Guinea, Francisco},
  journal = {Phys. Rev. Lett.},
  volume = {135},
  issue = {13},
  pages = {136503},
  numpages = {9},
  year = {2025},
  month = {Sep},
  publisher = {American Physical Society},
  doi = {10.1103/zfmh-rjzc},
  url = {https://link.aps.org/doi/10.1103/zfmh-rjzc}
}

@misc{guerci2024,
      title={Topological superconductivity from repulsive interactions in twisted {WSe$_2$}}, 
      author={Daniele Guerci and Daniel Kaplan and Julian Ingham and J. H. Pixley and Andrew J. Millis},
      year={2024},
      eprint={2408.16075},
      archivePrefix={arXiv},
      primaryClass={cond-mat.supr-con},
      url={https://arxiv.org/abs/2408.16075}, 
}

@article{chen2022excitonic,
  title={Excitonic insulator in a heterojunction moir{\'e} superlattice},
  author = {Chen, Dongxue and Lian, Zhen and Huang, Xiong and Su, Ying and Rashetnia, Mina and Ma, Lei and Yan, Li and Blei, Mark and Xiang, Li and Taniguchi, Takashi and Watanabe, Kenji and Tongay, Sefaattin and Smirnov, Dmitry and Wang, Zenghui and Zhang, Chuanwei and Cui, Yong-Tao and Shi, Su-Fei},
  journal={Nature Physics},
  volume={18},
  number={10},
  pages={1171--1176},
  year={2022},
  publisher={Nature Publishing Group UK London},
  url={https://doi.org/10.1038/s41567-022-01703-y}
}

@article{zhang2022correlated,
 author = {Zhang, Zuocheng and Regan, Emma C. and Wang, Danqing and Zhao, Wenyu and Wang, Shaoxin and Sayyad, Mohammed and Yumigeta, Kentaro and Watanabe, Kenji and Taniguchi, Takashi and Tongay, Sefaattin and Crommie, Michael and Zettl, Alex and Zaletel, Michael P. and Wang, Feng},
 issn = {1745-2481},
 journal = {Nat. Phys.},
 month = {August},
 number = {10},
 pages = {1214--1220},
 publisher = {Springer Science and Business Media LLC},
 title = {{Correlated} interlayer exciton insulator in heterostructures of monolayer $\mathrm{W}\mathrm{Se}_{2}$ and moir\'{e} $\mathrm{W}\mathrm{S}_{2}$ / $\mathrm{W}\mathrm{Se}_{2}$},
 url = {http://dx.doi.org/10.1038/s41567-022-01702-z},
 volume = {18},
 year = {2022}
}

@article{liu2022crossover,
 author = {Liu, Xiaomeng and Li, J. I. A. and Watanabe, Kenji and Taniguchi, Takashi and Hone, James and Halperin, Bertrand I. and Kim, Philip and Dean, Cory R.},
 issn = {1095-9203},
 journal = {Science},
 month = {January},
 number = {6577},
 pages = {205--209},
 publisher = {American Association for the Advancement of Science (AAAS)},
 title = {{Crossover} between strongly coupled and weakly coupled exciton superfluids},
 url = {http://dx.doi.org/10.1126/science.abg1110},
 volume = {375},
 year = {2022}
}

@article{Koshino_trigonal_2009,
  author = {Koshino, Mikito and McCann, Edward},
  title = {{Trigonal Warping and Berry's Phase {$N\pi$} in ABC-Stacked Multilayer Graphene}},
  journal = {Phys. Rev. B},
  volume = {80},
  pages = {165409},
  year = {2009},
  doi = {10.1103/PhysRevB.80.165409}
}

@article{Kumar_flat_2013,
  author = {Kumar, A. and Nandkishore, R.},
  title = {{Flat Bands with Berry Curvature in Multilayer Graphene}},
  journal = {Phys. Rev. B},
  volume = {87},
  pages = {241108},
  year = {2013},
  doi = {10.1103/PhysRevB.87.241108}
}

@article{Dong_anomalous_2024,
 author = {Dong, Junkai and Wang, Taige and Wang, Tianle and Soejima, Tomohiro and Zaletel, Michael P. and Vishwanath, Ashvin and Parker, Daniel E.},
 issn = {1079-7114},
 journal = {Phys. Rev. Lett.},
 month = {November},
 number = {20},
 pages = {206503},
 publisher = {American Physical Society (APS)},
 title = {{Anomalous} {Hall} {Crystals} in {Rhombohedral} {Multilayer} {Graphene}. {I}. {Interaction-Driven} {Chern} {Bands} and {Fractional} {Quantum} {Hall} {States} at {Zero} {Magnetic} {Field}},
 url = {http://dx.doi.org/10.1103/PhysRevLett.133.206503},
 volume = {133},
 year = {2024}
}

@article{Soejima_anomalous_2024,
 author = {Soejima, Tomohiro and Dong, Junkai and Wang, Taige and Wang, Tianle and Zaletel, Michael P. and Vishwanath, Ashvin and Parker, Daniel E.},
 issn = {2469-9969},
 journal = {Phys. Rev. B},
 month = {November},
 number = {20},
 pages = {205124},
 publisher = {American Physical Society (APS)},
 title = {{Anomalous} {Hall} crystals in rhombohedral multilayer graphene. {II}. {General} mechanism and a minimal model},
 url = {http://dx.doi.org/10.1103/PhysRevB.110.205124},
 volume = {110},
 year = {2024}
}

@article{Bardeen1957,
    title = {{Theory of Superconductivity}},
    year = {1957},
    journal = {Physical Review},
    author = {Bardeen, J. and Cooper, L. N. and Schrieffer, J. R.},
    number = {5},
    month = {12},
    pages = {1175--1204},
    volume = {108},
    publisher = {American Physical Society},
    url = {https://link.aps.org/doi/10.1103/PhysRev.108.1175},
    doi = {10.1103/PhysRev.108.1175},
    issn = {0031-899X}
}

@article{Anderson_morel,
 author = {Anderson, P. W. and Morel, P.},
 issn = {0031-9007},
 journal = {Phys. Rev. Lett.},
 month = {August},
 number = {4},
 pages = {136--138},
 publisher = {American Physical Society (APS)},
 title = {{Generalized} {Bardeen-Cooper-Schrieffer} {States} and {Aligned} {Orbital} {Angular} {Momentum} in the {Proposed} {Low-Temperature} {Phase} of {Liquid} {He}$^{3}$},
 url = {http://dx.doi.org/10.1103/PhysRevLett.5.136},
 volume = {5},
 year = {1960}
}

@article{Balian1963,
  title = {{Superconductivity with Pairs in a Relative $p$ Wave}},
  author = {Balian, R. and Werthamer, N. R.},
  journal = {Phys. Rev.},
  volume = {131},
  issue = {4},
  pages = {1553--1564},
  numpages = {0},
  year = {1963},
  month = {Aug},
  publisher = {American Physical Society},
  doi = {10.1103/PhysRev.131.1553},
  url = {https://link.aps.org/doi/10.1103/PhysRev.131.1553}
}

@article{ho2009ground,
 author = {Ho, L. H. and Micolich, A. P. and Hamilton, A. R. and Sushkov, O. P.},
 issn = {1550-235X},
 journal = {Phys. Rev. B},
 month = {October},
 number = {15},
 pages = {155412},
 publisher = {American Physical Society (APS)},
 title = {{Ground-plane} screening of {Coulomb} interactions in two-dimensional systems: {How} effectively can one two-dimensional system screen interactions in another},
 url = {http://dx.doi.org/10.1103/physrevb.80.155412},
 volume = {80},
 year = {2009}
}

@article{Olsen2016,
  title = {Simple Screened Hydrogen Model of Excitons in Two-Dimensional Materials},
  author = {Olsen, Thomas and Latini, Simone and Rasmussen, Filip and Thygesen, Kristian S.},
  journal = {Phys. Rev. Lett.},
  volume = {116},
  issue = {5},
  pages = {056401},
  numpages = {5},
  year = {2016},
  month = {Feb},
  publisher = {American Physical Society},
  doi = {10.1103/PhysRevLett.116.056401},
  url = {https://link.aps.org/doi/10.1103/PhysRevLett.116.056401}
}

@article{rohlfing2000electron,
  title = {Electron-hole excitations and optical spectra from first principles},
  author = {Rohlfing, Michael and Louie, Steven G.},
  journal = {Phys. Rev. B},
  volume = {62},
  issue = {8},
  pages = {4927--4944},
  numpages = {0},
  year = {2000},
  month = {Aug},
  publisher = {American Physical Society},
  doi = {10.1103/PhysRevB.62.4927},
  url = {https://link.aps.org/doi/10.1103/PhysRevB.62.4927}
}

@article{Anderson_Morel_2,
  title = {{Generalized Bardeen-Cooper-Schrieffer States and the Proposed Low-Temperature Phase of Liquid ${\mathrm{He}}^{3}$}},
  author = {Anderson, P. W. and Morel, P.},
  journal = {Phys. Rev.},
  volume = {123},
  issue = {6},
  pages = {1911--1934},
  numpages = {0},
  year = {1961},
  month = {Sep},
  publisher = {American Physical Society},
  doi = {10.1103/PhysRev.123.1911},
  url = {https://link.aps.org/doi/10.1103/PhysRev.123.1911}
}

@article{laturia2018dielectric,
 author = {Laturia, Akash and Van de Put, Maarten L. and Vandenberghe, William G.},
 issn = {2397-7132},
 journal = {npj 2D Materials and Applications},
 month = {March},
 number = {1},
 pages = {6},
 publisher = {Springer Science and Business Media LLC},
 title = {{Dielectric} properties of hexagonal boron nitride and transition metal dichalcogenides: from monolayer to bulk},
 url = {http://dx.doi.org/10.1038/s41699-018-0050-x},
 volume = {2},
 year = {2018}
}

@article{Dillenschneider_2008,
  title = {Exciton formation in graphene bilayer},
  author = {Dillenschneider, Raoul and Han, Jung Hoon},
  journal = {Phys. Rev. B},
  volume = {78},
  issue = {4},
  pages = {045401},
  numpages = {9},
  year = {2008},
  month = {Jul},
  publisher = {American Physical Society},
  doi = {10.1103/PhysRevB.78.045401},
  url = {https://link.aps.org/doi/10.1103/PhysRevB.78.045401}
}

@article{Ju_tunable_2017,
 author = {Ju, Long and Wang, Lei and Cao, Ting and Taniguchi, Takashi and Watanabe, Kenji and Louie, Steven G. and Rana, Farhan and Park, Jiwoong and Hone, James and Wang, Feng and McEuen, Paul L.},
 issn = {1095-9203},
 journal = {Science},
 month = {November},
 number = {6365},
 pages = {907--910},
 publisher = {American Association for the Advancement of Science (AAAS)},
 title = {{Tunable} excitons in bilayer graphene},
 url = {http://dx.doi.org/10.1126/science.aam9175},
 volume = {358},
 year = {2017}
}

@article{Quintela_tunable_2022,
 author = {Quintela, M. F. C. Martins and Peres, N. M. R.},
 issn = {2469-9969},
 journal = {Phys. Rev. B},
 month = {May},
 number = {20},
 pages = {205417},
 publisher = {American Physical Society (APS)},
 title = {{Tunable} excitons in rhombohedral trilayer graphene},
 url = {http://dx.doi.org/10.1103/PhysRevB.105.205417},
 volume = {105},
 year = {2022}
}

@article{Davenport_berry_2026,
 author = {Davenport, Henry and Schindler, Frank and Knolle, Johannes},
 issn = {2469-9969},
 journal = {Phys. Rev. B},
 month = {March},
 number = {11},
 pages = {115102},
 publisher = {American Physical Society (APS)},
 title = {{Berry} curvature of low-energy excitons in rhombohedral graphene},
 url = {http://dx.doi.org/10.1103/shrm-4swg},
 volume = {113},
 year = {2026}
}

@article{Scammell_2023,
  title = {Exciton condensation in biased bilayer graphene},
  author = {Scammell, Harley D. and Sushkov, Oleg P.},
  journal = {Phys. Rev. Res.},
  volume = {5},
  issue = {4},
  pages = {043176},
  numpages = {8},
  year = {2023},
  month = {Nov},
  publisher = {American Physical Society},
  doi = {10.1103/PhysRevResearch.5.043176},
  url = {https://link.aps.org/doi/10.1103/PhysRevResearch.5.043176}
}

@article{zverevich2026spintripletpairedwignercrystal,
 author = {Zverevich, Dmitry and Levchenko, Alex and Esterlis, Ilya},
 issn = {1079-7114},
 journal = {Phys. Rev. Lett.},
 month = {July},
 number = {3},
 pages = {036506},
 publisher = {American Physical Society (APS)},
 title = {{Spin-Triplet} {Paired} {Wigner} {Crystal} {Stabilized} by {Quantum} {Geometry}},
 url = {http://dx.doi.org/10.1103/ksxq-1l4q},
 volume = {137},
 year = {2026}
}

@article{Qi_observation_2026,
 author = {Qi, Ruishi and Li, Qize and Nie, Jiahui and Xia, Ruichen and Kim, Haleem and Lim, Hyungbin and Xie, Jingxu and Taniguchi, Takashi and Watanabe, Kenji and Crommie, Michael F. and MacDonald, Allan H. and Wang, Feng},
 issn = {1476-4687},
 journal = {Nature},
 month = {June},
 number = {8119},
 pages = {629--634},
 publisher = {Springer Science and Business Media LLC},
 title = {{Two-component} exciton condensates in an electron–hole bilayer},
 url = {http://dx.doi.org/10.1038/s41586-026-10636-y},
 volume = {654},
 year = {2026}
}

@article{Steinhoff_biexciton_2018,
 author = {Steinhoff, Alexander and Florian, Matthias and Singh, Akshay and Tran, Kha and Kolarczik, Mirco and Helmrich, Sophia and Achtstein, Alexander W. and Woggon, Ulrike and Owschimikow, Nina and Jahnke, Frank and Li, Xiaoqin},
 issn = {1745-2481},
 journal = {Nat. Phys.},
 month = {October},
 number = {12},
 pages = {1199--1204},
 publisher = {Springer Science and Business Media LLC},
 title = {{Biexciton} fine structure in monolayer transition metal dichalcogenides},
 url = {http://dx.doi.org/10.1038/s41567-018-0282-x},
 volume = {14},
 year = {2018}
}

@article{Feng_highly_2024,
 author = {Feng, Shun and Campbell, Aidan J. and Brotons-Gisbert, Mauro and Andres-Penares, Daniel and Baek, Hyeonjun and Taniguchi, Takashi and Watanabe, Kenji and Urbaszek, Bernhard and Gerber, Iann C. and Gerardot, Brian D.},
 issn = {2041-1723},
 journal = {Nat. Commun.},
 month = {May},
 number = {1},
 publisher = {Springer Science and Business Media LLC},
 title = {{Highly} tunable ground and excited state excitonic dipoles in multilayer 2{H-MoSe2}},
 url = {http://dx.doi.org/10.1038/s41467-024-48476-x},
 volume = {15},
 year = {2024}
}

@article{Hao_neutral_2017,
 author = {Hao, Kai and Specht, Judith F. and Nagler, Philipp and Xu, Lixiang and Tran, Kha and Singh, Akshay and Dass, Chandriker Kavir and Sch\"{u}ller, Christian and Korn, Tobias and Richter, Marten and Knorr, Andreas and Li, Xiaoqin and Moody, Galan},
 issn = {2041-1723},
 journal = {Nat. Commun.},
 month = {June},
 number = {1},
 pages = {15552},
 publisher = {Springer Science and Business Media LLC},
 title = {{Neutral} and charged inter-valley biexcitons in monolayer $\mathrm{Mo}\mathrm{Se}_{2}$},
 url = {http://dx.doi.org/10.1038/ncomms15552},
 volume = {8},
 year = {2017}
}

@article{Selig_ultrafast_2019,
 author = {Selig, Malte and Katsch, Florian and Schmidt, Robert and Michaelis de Vasconcellos, Steffen and Bratschitsch, Rudolf and Malic, Ermin and Knorr, Andreas},
 issn = {2643-1564},
 journal = {Phys. Rev. Research},
 month = {September},
 number = {2},
 pages = {022007},
 publisher = {American Physical Society (APS)},
 title = {{Ultrafast} dynamics in monolayer transition metal dichalcogenides: {Interplay} of dark excitons, phonons, and intervalley exchange},
 url = {http://dx.doi.org/10.1103/PhysRevResearch.1.022007},
 volume = {1},
 year = {2019}
}

@article{Ikegami_chiral_2013,
 author = {Ikegami, H. and Tsutsumi, Y. and Kono, K.},
 issn = {1095-9203},
 journal = {Science},
 month = {July},
 number = {6141},
 pages = {59--62},
 publisher = {American Association for the Advancement of Science (AAAS)},
 title = {{Chiral} {Symmetry} {Breaking} in {Superfluid}
3
{He-A}},
 url = {http://dx.doi.org/10.1126/science.1236509},
 volume = {341},
 year = {2013}
}

@article{PhysRevB.99.115439,
  title = {Spectrum of exciton states in monolayer transition metal dichalcogenides: Angular momentum and Landau levels},
  author = {Van der Donck, M. and Peeters, F. M.},
  journal = {Phys. Rev. B},
  volume = {99},
  issue = {11},
  pages = {115439},
  numpages = {9},
  year = {2019},
  month = {Mar},
  publisher = {American Physical Society},
  doi = {10.1103/PhysRevB.99.115439},
  url = {https://link.aps.org/doi/10.1103/PhysRevB.99.115439}
}

@article{huang2022excitons,
 author = {Huang, Di and Choi, Junho and Shih, Chih-Kang and Li, Xiaoqin},
 issn = {1748-3395},
 journal = {Nat. Nanotechnol.},
 month = {March},
 number = {3},
 pages = {227--238},
 publisher = {Springer Science and Business Media LLC},
 title = {{Excitons} in semiconductor moir\'{e} superlattices},
 url = {http://dx.doi.org/10.1038/s41565-021-01068-y},
 volume = {17},
 year = {2022}
}

@article{huang2023mott,
 author = {Huang, Tsung-Sheng and Chou, Yang-Zhi and Baldwin, Christopher L. and Wu, Fengcheng and Hafezi, Mohammad},
 issn = {2469-9969},
 journal = {Phys. Rev. B},
 month = {May},
 number = {19},
 pages = {195151},
 publisher = {American Physical Society (APS)},
 title = {{Mott-moir\'{e}} excitons},
 url = {http://dx.doi.org/10.1103/PhysRevB.107.195151},
 volume = {107},
 year = {2023}
}

@article{holbrook2026real,
 author = {Holbrook, Madisen and Ingham, Julian and Kaplan, Daniel and Holtzman, Luke N. and Bierman, Brenna and Hou, Bowen and Olsen, Nicholas and Nashabeh, Luca and Li, Yiliu and Liu, Song and Zhu, Xiaoyang and Qiu, Diana Y. and Rhodes, Daniel and Barmak, Katayun and Hone, James C. and Queiroz, Raquel and Pasupathy, Abhay N.},
 issn = {1745-2481},
 journal = {Nat. Phys.},
 month = {March},
 number = {5},
 pages = {680--685},
 publisher = {Springer Science and Business Media LLC},
 title = {{Real-space} imaging of the band topology of transition metal dichalcogenides},
 url = {http://dx.doi.org/10.1038/s41567-026-03197-4},
 volume = {22},
 year = {2026}
}

@article{PhysRevB.102.195403,
  title = {Theory of moir\'e localized excitons in transition metal dichalcogenide heterobilayers},
  author = {Ruiz-Tijerina, David A. and Soltero, Isaac and Mireles, Francisco},
  journal = {Phys. Rev. B},
  volume = {102},
  issue = {19},
  pages = {195403},
  numpages = {17},
  year = {2020},
  month = {Nov},
  publisher = {American Physical Society},
  doi = {10.1103/PhysRevB.102.195403},
  url = {https://link.aps.org/doi/10.1103/PhysRevB.102.195403}
}

@article{Fernandez_1997,
  title = {Spin Degree of Freedom in Two Dimensional Exciton Condensates},
  author = {Fern\'andez-Rossier, J. and Tejedor, C.},
  journal = {Phys. Rev. Lett.},
  volume = {78},
  issue = {25},
  pages = {4809--4812},
  numpages = {0},
  year = {1997},
  month = {Jun},
  publisher = {American Physical Society},
  doi = {10.1103/PhysRevLett.78.4809},
  url = {https://link.aps.org/doi/10.1103/PhysRevLett.78.4809}
}

@article{Wu_Macdonald2015,
  title = {Theory of two-dimensional spatially indirect equilibrium exciton condensates},
  author = {Wu, Feng-Cheng and Xue, Fei and MacDonald, A. H.},
  journal = {Phys. Rev. B},
  volume = {92},
  issue = {16},
  pages = {165121},
  numpages = {13},
  year = {2015},
  month = {Oct},
  publisher = {American Physical Society},
  doi = {10.1103/PhysRevB.92.165121},
  url = {https://link.aps.org/doi/10.1103/PhysRevB.92.165121}
}

@article{Dinh_Dery2025,
  title = {Landau-level composition of bound exciton states in magnetic field},
  author = {Van Tuan, Dinh and Dery, Hanan},
  journal = {Phys. Rev. B},
  volume = {112},
  issue = {8},
  pages = {085305},
  numpages = {20},
  year = {2025},
  month = {Aug},
  publisher = {American Physical Society},
  doi = {10.1103/5msh-zgy9},
  url = {https://link.aps.org/doi/10.1103/5msh-zgy9}
}

@article{Quintela_theoretical_2022,
 author = {Quintela, Maurício F. C. Martins and Henriques, João C. G. and Tenório, Luiz G. M. and Peres, Nuno M. R.},
 issn = {1521-3951},
 journal = {physica status solidi (b)},
 month = {April},
 number = {7},
 publisher = {Wiley},
 title = {{Theoretical} {Methods} for {Excitonic} {Physics} in 2{D} {Materials}},
 url = {http://dx.doi.org/10.1002/pssb.202200097},
 volume = {259},
 year = {2022}
}

@article{Liu_optical_2025,
 author = {Liu, Xiaoling and Leisgang, Nadine and Dolgirev, Pavel E. and Zibrov, Alexander A. and Sung, Jiho and Wang, Jue and Taniguchi, Takashi and Watanabe, Kenji and Walther, Valentin and Park, Hongkun and Demler, Eugene and Kim, Philip and Lukin, Mikhail D.},
 issn = {1745-2481},
 journal = {Nat. Phys.},
 month = {August},
 number = {10},
 pages = {1563--1569},
 publisher = {Springer Science and Business Media LLC},
 title = {{Optical} signatures of interlayer electron coherence in a bilayer semiconductor},
 url = {http://dx.doi.org/10.1038/s41567-025-02971-0},
 volume = {21},
 year = {2025}
}

@article{Moon_exciton_2025,
 author = {Moon, Byoung Hee and Mondal, Ashok and Efimkin, Dmitry K. and Lee, Young Hee},
 issn = {2522-5820},
 journal = {Nat. Rev. Phys.},
 month = {June},
 number = {7},
 pages = {388--401},
 publisher = {Springer Science and Business Media LLC},
 title = {{Exciton} condensate in van der {Waals} layered materials},
 url = {http://dx.doi.org/10.1038/s42254-025-00834-4},
 volume = {7},
 year = {2025}
}

@article{Kwon_defect-bound_2026,
 author = {Kwon, SeongJin and Jin, Kyung-Hwan and Han, Jong Eun and Lee, Siwon and Won, ChoongJae and Cheong, Sang-Wook and Yeom, Han Woong},
 issn = {1748-3395},
 journal = {Nat. Nanotechnol.},
 month = {September},
 publisher = {Springer Science and Business Media LLC},
 title = {{Defect-bound} states of exciton condensate as an analogue of the {Yu}–{Shiba}–{Rusinov} state},
 url = {http://dx.doi.org/10.1038/s41565-026-02267-1},
 year = {2026}
}

@article{Panigrahi_tomographic_2026,
 author = {Panigrahi, Archisman and Poliakov, Vladislav and Levitov, Leonid},
 issn = {1091-6490},
 journal = {Proceedings of the National Academy of Sciences},
 month = {July},
 number = {27},
 pages = {e2534730123},
 publisher = {National Academy of Sciences},
 title = {{Tomographic} imaging of superconducting order using particle–hole interference},
 url = {http://dx.doi.org/10.1073/pnas.2534730123},
 volume = {123},
 year = {2026}
}

\onecolumngrid
\appendix
\setcounter{figure}{0}
\renewcommand{\thefigure}{\thesection\arabic{figure}}

\numberwithin{equation}{section}

\setcounter{equation}{0}
\renewcommand{\theequation}{A\arabic{equation}}
\section*{Appendix A: Derivation of the Bethe-Salpeter Equation}

Consider the valence band and the conduction band, with dispersion $\varepsilon_v(\vec k)$ and $\varepsilon_c(\vec k)$, respectively. The chemical potential $\mu$ is somewhere within the gap, i.e. $\forall \vec k$, $\varepsilon_v(\vec k) < \mu < \varepsilon_c(\vec k)$. Note that all the momenta are measured from $K$-point in the BZ. For simplicity, we drop the electron-electron interactions within the same band, and only retain the projected inter-band interaction
\begin{equation}
    H = \sum_{\vec k} (\varepsilon_c(\vec k) - \mu) c^\dagger_{c \vec{k}} c_{c \vec{k}}+ (\varepsilon_v(\vec k) - \mu) c^\dagger_{v \vec{k}} c_{v \vec{k}} + \frac{1}{A} \sum_{\vec k, \vec k', \vec q} V(\vec q) c^\dagger_{c(\vec k + \vec q)} c^\dagger_{v(\vec k' - \vec q)} c_{v(\vec k')} c_{c \vec{k}} \Lambda_{cc}^{\vec k + \vec q, \vec k} \Lambda_{vv}^{\vec k' - \vec q, \vec k'}
\end{equation}

Now we define a hole operator, $d^\dagger_{\vec k} = c_{v{(-\vec k)}}$ (more precisely, $d^\dagger_{(\vec K + \vec k)} = c_{v(\vec K -\vec k)}$). Also, henceforth we denote the conduction band annihilation operator $c_{c \vec{k}}$ as just $ c_{\vec k}$. We can write the Hamiltonian in terms of these new $d$ operators (and drop overall constants and switch dummy variable $\vec k' \rightarrow (-\vec k' + \vec q)$),
\begin{equation}\label{eq:effective-Hamiltonian}
    H' = \sum_{\vec k} (\varepsilon_c(\vec k) - \mu) c^\dagger_{ \vec{k}} c_{ \vec{k}} - (\varepsilon_v(-\vec k) - \mu) d^\dagger_{\vec k} d_{\vec k} - \frac{1}{A} \sum_{\vec k, \vec k', \vec q} V(\vec q) c^\dagger_{\vec k + \vec q} d^\dagger_{\vec k'-\vec q} d_{\vec k'} c_{ \vec{k}} \Lambda_{cc}^{\vec k + \vec q, \vec k} \Lambda_{vv}^{-\vec k', -\vec k'+\vec q}
\end{equation}
We define $\ket{\text{val}}$ to be the many body state where the upper band is completely empty and the lower band is completely filled. In other words $c_{\vec k} \ket{\text{val}} =0 = c_{v\vec k}^\dagger \ket{\text{val}} = d_{\vec k} \ket{\text{val}}$.

Then, the energy of the state without any exciton (band insulator) is
\begin{equation}\label{eq:ref-energy}
    H'{\ket{\rm val}} = 0.
\end{equation}

We want to solve a single exciton problem with zero center of mass momentum, instead of solving the many-body problem. Therefore, we restrict the interaction to the case where $\vec k' = -\vec k$.
We define a single exciton state, $\ket{1X} = \sum_{\vec k} A_{\vec k} {c}^\dagger_{\vec k} d^\dagger_{-\vec k} \ket{\text{val}}$ (more precisely, it is $\ket{1X} = \sum_{\vec k} A_{\vec k} {c}^\dagger_{\vec K + \vec k} d^\dagger_{\vec K -\vec k} \ket{\text{val}}$).

When we apply $H'$ on this state, only the terms with $\vec k' = -\vec k$ survive.

We obtain,
\begin{equation}
    H' \ket{1X} =  \sum_{\vec k} (\varepsilon_c(\vec k) - \varepsilon_v(\vec k))A_{\vec k} {c}^\dagger_{\vec k} d^\dagger_{-\vec k} \ket{\text{val}} -\frac{1}{A} \sum_{\vec k , \vec q} \left[V(\vec q) \Lambda_{cc}^{\vec k + \vec q, \vec k} \Lambda_{vv}^{\vec k, \vec k+\vec q}\right] A_{\vec k} {c}^\dagger_{\vec k +\vec q} d^\dagger_{-\vec k-\vec q} 
    \ket{\text{val}}
\end{equation}

Let us study the interaction term, where we get rid of the dummy variable $\vec q$ in favor of a new dummy variable $\vec k''=\vec k + \vec q$,
\begin{equation}
\begin{aligned}
    H_{\rm int} \ket{\rm val} &= -\frac{1}{A} \sum_{\vec k , \vec k''} \left[V(\vec k''-\vec k) \Lambda_{cc}^{\vec k'', \vec k} \Lambda_{vv}^{\vec k, \vec k''}\right] A_{\vec k} {c}^\dagger_{\vec k''} d^\dagger_{-\vec k''} 
    \ket{\text{val}}\\
    &= -\frac{1}{A} \sum_{\vec k , \vec k''} \left[V(\vec k-\vec k'') \Lambda_{cc}^{\vec k, \vec k''} \Lambda_{vv}^{\vec k'', \vec k}\right] A_{\vec k''} {c}^\dagger_{\vec k} d^\dagger_{-\vec k} 
    \ket{\text{val}} \text{ (swapping dummy variables $\vec k \leftrightarrow \vec k''$)}
\end{aligned}
\end{equation}

Using this, we obtain the eigenvalue problem,
\begin{equation}\label{eq:general-single-particle-problem}
   (\varepsilon_c(\vec k) - \varepsilon_v(\vec k)) A_{\vec k}  -\frac{1}{A} \sum_{\vec k'} \left[V(\vec k-\vec k') \Lambda_{cc}^{\vec k, \vec k'} \Lambda_{vv}^{\vec k', \vec k}\right] A_{\vec k'} = E A_{\vec k}
\end{equation}

For bound state to exist, we must have $E' < E'_{\ket{\rm val}}=0$ (we have chosen Hamiltonian $H'$ such that the energy $E'_{\ket{\rm val}}$ of the state without excitons is exactly zero. See Eqs.~\eqref{eq:effective-Hamiltonian} and~\eqref{eq:ref-energy}).

\subsection{Simplification in the presence of rotational invariance: Angular harmonics decomposition}
Let's assume that the problem has rotational invariance.
Due to rotational invariance, we can look into a particlar angular momentum channel, $A_{\vec k} = \sum_l \psi_l(k) e^{i l \theta}$. Then,
\begin{equation}\label{eq:rotational-invariance-simplification}
    \begin{aligned}
        &-\frac{1}{A} \sum_{\vec k , \vec k''} \left[V(\vec k-\vec k'') \Lambda_{cc}^{\vec k, \vec k''}\Lambda_{vv}^{\vec k'', \vec k}\right] A_{\vec k''} \\
        &= -\frac{1}{L^2} \sum_l \int_{0}^\infty \frac{k'' dk''}{(2\pi)^2} \int_0^{2\pi} d\chi \left[V(\sqrt{k^2 + k''^2 - 2 k k'' \cos\chi}) \Lambda_{cc}( k, k'';\chi) \Lambda_{vv}( k'', k,-\chi)\right] \psi_l(k'') e^{i l (\theta + \chi)}\\
        &= -\frac{1}{L} \sum_{l,k''} k'' \psi_l(k'') e^{i l \theta} \left[\int_0^{2\pi} \frac{d\chi}{2\pi} \left[V(\sqrt{k^2 + k''^2 - 2 k k'' \cos\chi}) \Lambda_{cc}( k, k'';\chi) \Lambda_{vv}( k'', k,-\chi)\right] e^{i l\chi}\right]\\
        &= -\frac{1}{L} \sum_{l,k''} k'' \psi_l(k'') e^{i l \theta} {V_l}(k,k'';\Lambda)
    \end{aligned}
\end{equation}
where
$\Lambda_{cc/vv}(k,k'';\xi) = \Lambda_{cc/vv}^{\vec k,\vec k''}$ such that the angle between $\vec k$ and $\vec k''$ is $\chi$, and
\begin{equation}\label{eq:V-l-k-k'}
    {V}_l(k,k'';\Lambda) = \left[\int_0^{2\pi} \frac{d\chi}{2\pi} \left[V(\sqrt{k^2 + k''^2 - 2 k k'' \cos\chi}) \Lambda_{cc}( k, k'';\chi) \Lambda_{vv}( k'', k,-\chi)\right] e^{i l\chi}\right]
\end{equation}
Note: In the main text, we defined the quantity $\tilde{V}_l(k,k'')$, which is the non-dimensionalized version of ${V}_l(k,k'';\Lambda)$ defined here.
We can use this to separate the channel $l$ in Eq.~\eqref{eq:general-single-particle-problem} and obtain,
\begin{equation}\label{eq:channel-l-working-equation}
    \left[\varepsilon_c( k) - \varepsilon_v(k)\right] \psi_l( k) -\frac{\Delta k''}{2\pi} \sum_{k''} k''  {V_l}(k,k'';\Lambda) \psi_l(k'') = E \psi_l(k)
\end{equation}
where $ \Delta k'' = \frac{2\pi}{L}$ is the radial grid spacing.
We can solve this eigenvalue problem numerically and compare the lowest energy eigenvalue in each channel.

In other words, this equation is telling us, bound state energy $E' \sim (E_{\rm bandgap} + E_{\rm kinetic} - |V|)$.

\setcounter{equation}{0}
\renewcommand{\theequation}{B\arabic{equation}}
\section*{Appendix B: Analysis of chiral excitonic condensate}
We start with the Hamiltonian $H'$ defined in Eq.~\eqref{eq:effective-Hamiltonian} and do mean-field in the exciton channel (dropping all the overall constants),
\begin{equation}
\begin{aligned}
    H'_{MF} = &\sum_{\vec k} (\varepsilon_c(\vec k) - \mu) c^\dagger_{\vec{k}} c_{ \vec{k}} - (\varepsilon_v(\vec k) - \mu) d^\dagger_{-\vec k} d_{-\vec k} \\ &- \frac{1}{A} \sum_{\vec k, \vec k', \vec q} \left[V(\vec q)\Lambda_{cc}^{\vec k + \vec q, \vec k} \Lambda_{vv}^{-\vec k', -\vec k'+\vec q}\right]  \left(\expval{ c^\dagger_{\vec k + \vec q}  d^\dagger_{\vec k'-\vec q}} d_{\vec k'} c_{ \vec{k}} + c^\dagger_{\vec k + \vec q}  d^\dagger_{\vec k'-\vec q} \expval{d_{\vec k'} c_{ \vec{k}}} \right).
\end{aligned}
\end{equation}

Now we define the excitonic gap function,
\begin{equation}\label{eq:gap-eqn-prelim}
    \Delta_{\vec k} = - \frac{1}{A} \sum_{\vec q} \left[V(\vec q)\Lambda_{cc}^{\vec k + \vec q, \vec k} \Lambda_{vv}^{-\vec k', -\vec k'+\vec q}\right] \expval{d_{-\vec k -\vec q}c_{\vec k + \vec q}}
\end{equation}
Here $V(q) >0$ between bare electrons repulsive, but compared to BdG gap equation in superconductivity, there is an overall negative sign, which is why the excitonic gap equation is satisfied for repulsive interaction between electrons.

This lets us write,
\begin{equation}
        H'_{MF} = \sum_{\vec k} (\varepsilon_c(\vec k) - \mu) c^\dagger_{c \vec{k}} c_{ \vec{k}} - (\varepsilon_v(\vec k) - \mu) d^\dagger_{-\vec k} d_{-\vec k} + \sum_{\vec k, \vec k'} (\Delta^*_{\vec k} d_{-\vec k}c_{ \vec{k}} + \Delta_{\vec k}c^\dagger_{\vec k}  d^\dagger_{-\vec k} )
\end{equation}

Now we introduce a Bogoliubov transformation,
\begin{equation}
    \begin{aligned}
        {\gamma_1}_{\vec k} = u_{\vec k}c_{\vec k} + v_{\vec k} d^\dagger_{-\vec k}\\
        {\gamma_2}_{-\vec k} = -v_{\vec k}c^\dagger_{\vec k} + u_{\vec k} d_{-\vec k}\\
    \end{aligned}
\end{equation}

We invert the Bogoliubov transformation, which allows us to write $H'_{MF}$ in terms of the quasiparticle operators,
\begin{equation}
\begin{aligned}
    H'_{MF} &= \sum_{\vec k} \left[\sqrt{|\Delta_{\vec k}|^2 + \left(\frac{\varepsilon_c(\vec k) - \varepsilon_v(\vec k)}{2}\right)^2} + \left(\frac{\varepsilon_c(\vec k) + \varepsilon_v(\vec k)}{2} - \mu \right)\right] {\gamma_1}^\dagger_{\vec k} {\gamma_1}_{\vec k} 
    \\ &+ \sum_{\vec k} \left[\sqrt{|\Delta_{\vec k}|^2 + \left(\frac{\varepsilon_c(\vec k) - \varepsilon_v(\vec k)}{2}\right)^2 } - \left(\frac{\varepsilon_c(\vec k) + \varepsilon_v(\vec k)}{2} - \mu\right)\right] {\gamma_2}^\dagger_{-\vec k} {\gamma_2}_{-\vec k}
\end{aligned}
\end{equation}
We identify the excitation energies with $E_1(\vec k)$ and $E_2(\vec k)$, as mentioned in the main text.

After substituting the $c, d$ operators in terms of $\gamma$ operators, the gap equation in Eq.~\eqref{eq:gap-eqn-prelim} takes the form
\begin{equation}
    \Delta_{\vec k} = \frac{1}{A} \sum_{\vec q}  \left[V(\vec q)\Lambda_{cc}^{\vec k + \vec q, \vec k} \Lambda_{vv}^{\vec k, \vec k+\vec q}\right] \frac{\Delta_{\vec k + \vec q}}{(E_1(\vec k+\vec q) + E_2({\vec k+\vec q}))} \left[1 - f(E_1(\vec k+\vec q)) -f(E_2 (\vec k+\vec q)) \right]
\end{equation}

Near $T_c$ we consider the linearized gap equation,
\begin{equation}
    \Delta_{\vec k} = \frac{1}{A} \sum_{\vec q}  \left[V(\vec q)\Lambda_{cc}^{\vec k + \vec q, \vec k} \Lambda_{vv}^{\vec k, \vec k+\vec q}\right]
    \frac{\Delta_{\vec k + \vec q}}{\varepsilon_c(\vec k + \vec q) - \varepsilon_v(\vec k + \vec q)} \tanh\left[\frac{\varepsilon_c( \vec k + \vec q) - \varepsilon_v(\vec k + \vec q)}{4 T} \right]
\end{equation}

If the system has cylindrical symmetry, we can into the $l$-th angular momentum channel, $\Delta_{\vec k} \sim \Delta_l e^{i l \theta}$, which satisfies the linearized gap equation,
\begin{equation}
    \Delta_{l} (k) = \frac{1}{L} \sum_{k'' >0} k'' V_{l} (k,k'';\Lambda) \frac{\tanh\left(\frac{\varepsilon_c( k'') - \varepsilon_v(k'')}{4 k_B T} \right)}{\varepsilon_c( k'') - \varepsilon_v(k'')} \Delta_l(k'')
\end{equation}
where $V_l(k,k'';\Lambda)$ has been defined before in Eq.~\eqref{eq:V-l-k-k'}.

We can cast this as an eigenvalue problem, and look into the largest eigenvalue of the kernel in the right hand side at a particular temperature. The angular momentum channel with the largest eigenvalue is the leading instability.

In the equation above, the kernel $\frac{1}{L} \sum_{k'' >0} k'' V_{l} (k,k'';\Lambda) \frac{\tanh\left(\frac{\varepsilon_c( k'') - \varepsilon_v(k'')}{4 k_B T} \right)}{\varepsilon_c( k'') - \varepsilon_v(k'')} $ is not Hermitian.
In general, for equations of the form,
\begin{equation}
    \Delta_l(k) = \sum_{k''} A(k,k'') F(k'') \Delta_l(k'')
\end{equation}
we can define the auxiliary function, $\phi(k) = \Delta_l(k) \sqrt{F(k)}$. Then, the equation above can be rewritten as,
\begin{equation}
    \phi_l(k) = \sum_{k''} A(k,k'') \sqrt{F(k) F(k')} \phi_l(k'').
\end{equation}
Here, the kernel becomes Hermitian.

Note: It can be shown that $V_l(k,k'';\Lambda)$ is real, i.e. $V_l(k,k'';\Lambda) = V_l(k,k'';\Lambda)^*$, and it is symmetric under interchanging $k\leftrightarrow k'$, i.e. $V_l(k,k'';\Lambda) = V_l(k'',k;\Lambda)$.

\setcounter{equation}{0}
\renewcommand{\theequation}{C\arabic{equation}}
\section*{Appendix C: Comparison between gap equation at zero temperature and the single-exciton bound state equation}

We may ask, does the existence of the bound state guarantee a solution of the linearized gap equation? The answer is affirmative, and this argument is based on a similar argument in Ref.~\cite{Jerome_Excitonic_1967}.

At zero temperature, the gap equation assumes the form,
\begin{equation}
    \Delta_{\vec k} = \frac{1}{A} \sum_{\vec q}  \left[V(\vec q)\Lambda_{cc}^{\vec k + \vec q, \vec k} \Lambda_{vv}^{\vec k, \vec k+\vec q}\right] \frac{\Delta_{\vec k + \vec q}}{(E_1(\vec k) + E_2({\vec k}))}
\end{equation}
Let $\phi_{\vec k} = \frac{\Delta_{\vec k}}{E_1(\vec k) + E_2(\vec k)}$. Then, the gap equation can be rewritten as,
\begin{equation}
    \left[\sqrt{4|\Delta_{\vec k}|^2 + ({\varepsilon_c(\vec k) - \varepsilon_v(\vec k)})^2} \right] \phi_{\vec k} - \frac{1}{A} \sum_{\vec q}  \left[V(\vec q)\Lambda_{cc}^{\vec k + \vec q, \vec k} \Lambda_{vv}^{\vec k, \vec k+\vec q}\right] \phi_{\vec k + \vec q} = 0.
\end{equation}

Compare it with the Schrodinger equation for a single exciton,
\begin{equation}
   (\varepsilon_c(\vec k) - \varepsilon_v(\vec k)) A_{\vec k}  -\frac{1}{A} \sum_{\vec k''} \left[V(\vec q)\Lambda_{cc}^{\vec k + \vec q, \vec k} \Lambda_{vv}^{\vec k, \vec k+\vec q}\right] A_{\vec k'} = E' A_{\vec k}
\end{equation}

Comparing between the two, we find that at zero temperature whenever a single exciton bound state forms (which means eigenvalue $E' < 0$), the excitonic order parameter has a non-trivial solution, $|\Delta_{\vec k}| > 0$.

\setcounter{equation}{0}
\renewcommand{\theequation}{D\arabic{equation}}

\section*{Appendix D: Bandstructure of rhombohedral tetralayer graphene}
We write down a six orbital band model of rhombohedral tetralayer graphene in the ordered basis $(A_1, B_1, A_2, B_2, A_3, B_3, A_4, B_4)$ \cite{MacDonald_ABCtrilayer_2010},
\begin{equation}
H(\mathbf{k})=
\begin{pmatrix}
U/2 & v_0\pi^\dagger & -v_4\pi^\dagger & v_3\pi & 0 & \gamma_2 & 0 & 0\\
v_0\pi & U/2 & \gamma_1 & -v_4\pi^\dagger & 0 & 0 & 0 & 0\\
-v_4\pi & \gamma_1 & U/6 & v_0\pi^\dagger & -v_4\pi^\dagger & v_3\pi & 0 & \gamma_2\\
v_3\pi^\dagger & -v_4\pi & v_0\pi & U/6 & \gamma_1 & -v_4\pi^\dagger & 0 & 0\\
0 & 0 & -v_4\pi & \gamma_1 & -U/6 & v_0\pi^\dagger & -v_4\pi^\dagger & v_3\pi\\
\gamma_2 & 0 & v_3\pi^\dagger & -v_4\pi & v_0\pi & -U/6 & \gamma_1 & -v_4\pi^\dagger\\
0 & 0 & 0 & 0 & -v_4\pi & \gamma_1 & -U/2 & v_0\pi^\dagger\\
0 & 0 & \gamma_2 & 0 & v_3\pi^\dagger & -v_4\pi & v_0\pi & -U/2
\end{pmatrix}
\end{equation}
with $\pi = k_x + i k_y$, and $v_i = \frac{\sqrt{3}}{2} a \gamma_i$ for $i=1,2,3,4$. Here $a = 2.46\text{\AA}$ is the lattice constant, and for hopping elements $\gamma_i$, we use the Slonczewski-Weiss-McClure (SWMc) tight-binding parameters, $\gamma_0 = 3.16$ eV, $\gamma_1 = 0.38$ eV, $\gamma_2 = -0.015$ eV, $\gamma_3 = 0.29$ eV and $\gamma_4 = 0.141$ eV.

To extract the contribution of a particular angular momentum channel in an energy eigenstate (Fig.~\ref{fig:graphene_tetralayer}(d) in main text), we compute the overlap of the ground state wavefunction with an angular momentum wavefunction $e^{i l \theta}$.

\section*{Appendix E: Spectrum of two-dimensional H-atom under magnetic field}

\setcounter{equation}{0}
\renewcommand{\theequation}{E\arabic{equation}}

\counterwithin*{figure}{part}

\stepcounter{part}

\renewcommand{\thefigure}{D.\arabic{figure}}

\begin{figure}[h]
    \centering
    \includegraphics[width=0.6\linewidth]{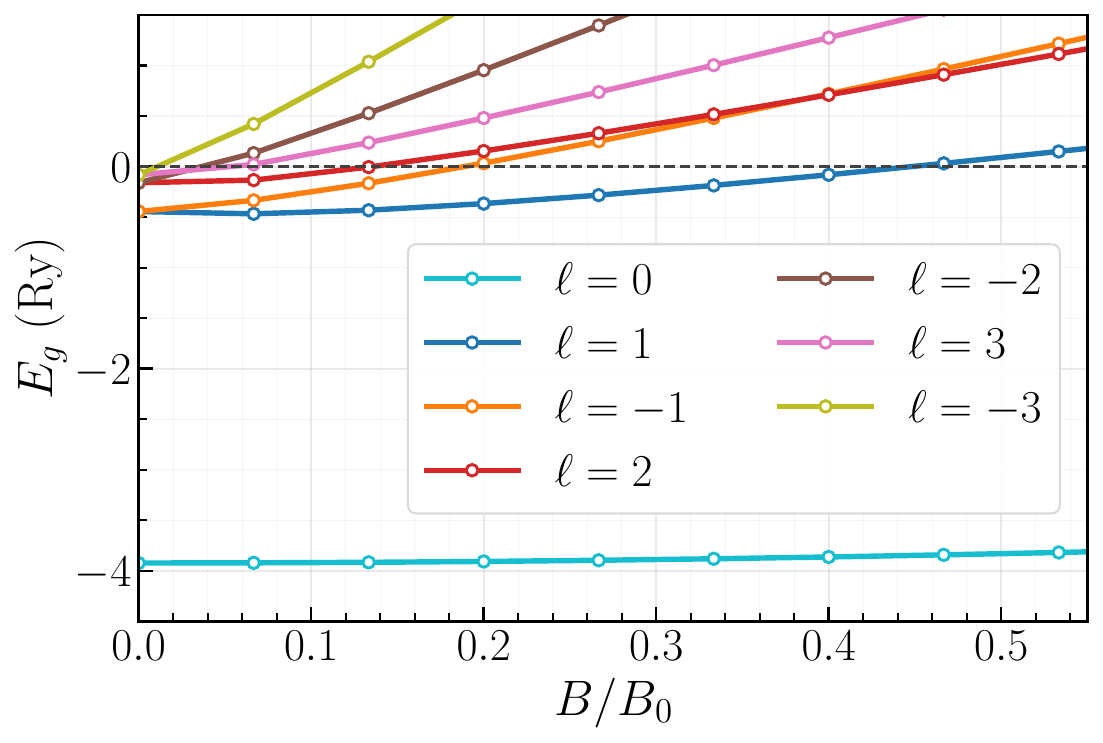}
    \caption{The dependence of ground state energy on magnetic field for different angular momentum channels of the two-dimensional hydrogen atom. The ground state energy is plotted in the units of Rydberg ($= 13.6 eV$), and the magnetic is expressed in terms of the atomic units, $B_0 = \frac{e^3 m^2}{16 \pi^2 \epsilon_0^2 \hbar^3} = 2.34 \times 10^5 $T. The ground state energy of $s$-wave at zero magnetic field is $-4$ Ry (Ref.~\cite{Yang_analytic_1991}). In the numerics, it deviated slightly from the analytical value due to finite grid size.}
    \label{fig:H-atom-B-field-dependence}
\end{figure}

We now solve the hydrogenic problem in two-dimension \cite{Olsen2016}, with the Berry phase's role replaced by a uniform magnetic field and show that the $s$-wave exciton is always the ground state. Consider the Hamiltonian,
\begin{equation}
H = \frac{(\vec p_{_{\rm 2D}} + e \vec A(r))^2}{2 m} - \frac{e^2}{4\pi \epsilon_0 r_{_{\rm 2D}}},
\end{equation}
with $\vec A = \frac{B}{2}(-y,x,0)$. We numerically diagonalize the Hamiltonian and find that while the energy of $p+ip$ state initially decreases at a small magnetic field, it can never overtake the $s$-wave as the ground state, as we show in the appendix, Fig.~\ref{fig:H-atom-B-field-dependence}.

\section*{Appendix F: Excitons under Yukawa interaction}

\setcounter{equation}{0}
\renewcommand{\theequation}{F\arabic{equation}}
\renewcommand{\thefigure}{E.\arabic{figure}}

Here we consider excitons in tetralayer graphene forming under Yukawa interaction, $V(k) = \frac{2 \pi e^2}{\epsilon_0 \epsilon_r (k + k_0)}$, which can be considered as an interpolant between the Coulomb interaction and the momentum independent Hubbard interaction. Here we take $k_0 = 0.02/$\AA, which is the typical distance between the Fermi pockets generated by trigonal warping.
We find that the phase diagram is qualitatively similar to the phase diagram with a gate-screened Coulomb interaction.
\begin{figure}[h]
    \centering
    \includegraphics[width=0.5\linewidth]{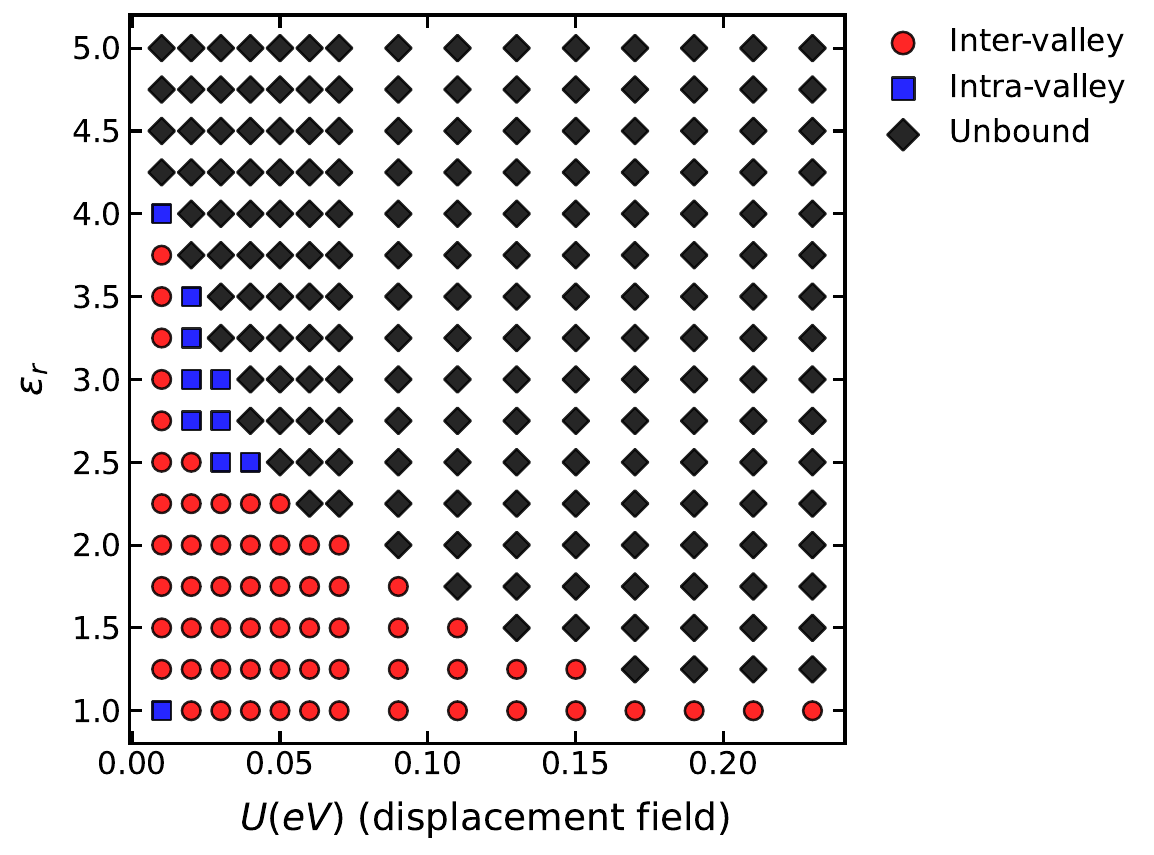}
    \caption{Phase diagram of excitons in tetralayer graphene under Yukawa interaction. See appendix F for details.}
    \label{fig:phase-diagram-Yukawa}
\end{figure}

\section*{Appendix G: Derivation of free-energy functional of excitons}

Here we perform a calculation inspired by the one in Ref.~\cite{Wu_Macdonald2015,Fernandez_1997} to compute the free energy functional to study the valley-polarization of excitonic condensates. For simplicity of the calculation, we will drop the form factors, and use bare Coulomb interaction to illustrate the effect.

The mean-field excitonic ground state has the BCS-like wavefunction
\begin{equation}
\ket{\Psi} = \prod_{k} (u_k + v_k c_k^\dagger d_{-k}^\dagger) |\text{vac}\rangle.
\end{equation}

Here $c_k = c_{c,k}$ and $d^\dagger_{-k} = c_{v,k}$. For simplicity, we define $a_{c, k} = c_k$ and $a_{v, k} = c_{v,k}$. We define the density matrix, $\rho_{ij} = \bra \Psi a_{i}^\dagger a_j\ket\Psi$.
The wavefunction can be rewritten as,
\begin{equation}
\ket{\Psi} = e^{\sum \lambda_{vc} a_c^\dagger a_v} |\text{vac}\rangle,
\end{equation}
where the matrix $\lambda_{vc}$ contains a set of complex parameters corresponding to coupling between different kinds of electron and hole operators, to create different types of excitons. Note, when we write $\lambda_{vc}$ to describe its elements, the subscript really carries the valley index and the momentum index, i.e., $\lambda_{[v,k_1],[c,k_2]}$.

If there was only a single species of exciton, then we could write,
\begin{equation}
\bra \Psi a_c^\dagger a_c \ket \Psi = \frac{|\lambda_i|^2}{1 + |\lambda_i|^2}, \quad \bra \Psi a_v^\dagger a_v \ket \Psi = \frac{1}{1 + |\lambda_i|^2}, \quad \bra \Psi a_v^\dagger a_c \ket \Psi= \frac{\lambda_i}{1 + |\lambda_i|^2}.
\end{equation}

Generalizing this relation, we obtain,
\begin{equation} \begin{aligned}
\rho_{vv'} &= \left[(1 + \lambda \lambda^\dagger)^{-1}\right]_{vv'} \approx \delta_{vv'} - (\lambda \lambda^\dagger)_{vv'} + (\lambda \lambda^\dagger \lambda \lambda^\dagger)_{vv'} \\
\rho_{cc'} &= \left[\lambda^\dagger (1 + \lambda \lambda^\dagger)^{-1} \lambda\right]_{cc'} = (\lambda^\dagger \lambda)_{cc'} - (\lambda^\dagger \lambda \lambda^\dagger \lambda)_{cc'} \\
\rho_{vc} &= \left[(1 + \lambda \lambda^\dagger)^{-1} \lambda\right]_{vc} = \lambda_{vc} - (\lambda \lambda^\dagger \lambda)_{vc}.
 \end{aligned} \end{equation}

Let $\lambda = \Lambda + \frac{1}{2} \Lambda \Lambda^\dagger \Lambda$. We will shortly show that $\Lambda$ rather than $\lambda$ is the natural expansion parameter in the limit of small excitonic density.

Then $\rho_{vv'}$ can be written in terms of $\Lambda$ as,
\begin{equation} \begin{aligned}
\rho_{vv'} &= \delta_{vv'} - \left(\Lambda + \frac{1}{2}\Lambda\Lambda^\dagger\Lambda\right)\left(\Lambda^\dagger + \frac{1}{2}\Lambda^\dagger\Lambda\Lambda^\dagger\right) + \left[\left(\Lambda + \frac{1}{2}\Lambda\Lambda^\dagger\Lambda\right)\left(\Lambda^\dagger + \frac{1}{2}\Lambda^\dagger\Lambda\Lambda^\dagger\right)\right]^2 \\
&= \delta_{vv'} - \Lambda\Lambda^\dagger - \Lambda\Lambda^\dagger\Lambda\Lambda^\dagger + \Lambda\Lambda^\dagger\Lambda\Lambda^\dagger + \mathcal{O}(\Lambda^6) \\
&\approx \delta_{vv'} - (\Lambda\Lambda^\dagger)_{vv'}
 \end{aligned} \end{equation}

Similarly, we obtain,
\begin{equation} \begin{aligned}
\rho_{cc'} &= (\Lambda^\dagger \Lambda)_{cc'} \\
\rho_{vc} &= \left(\Lambda - \frac{1}{2}\Lambda\Lambda^\dagger\Lambda\right)_{vc}.
 \end{aligned} \end{equation}

The density of excitons is identical to the density of (virtually excited) conduction electrons,
\begin{equation}
N_{\text{ex}} = \sum \rho_{cc} = \text{Tr}(\Lambda^\dagger \Lambda).
\end{equation}

We utilize a general excitonic Hamiltonian,
\begin{equation}
H = \sum_\alpha \varepsilon_\alpha a_\alpha^\dagger a_\alpha + \frac{1}{2}\sum_{\alpha\beta\gamma\delta} W_{\alpha\beta\gamma\delta} a_\alpha^\dagger a_\beta^\dagger a_\gamma a_\delta.
\end{equation}

It can be shown that $\rho^2 - \rho \sim \mathcal{O}(\Lambda^5)$, so $\Lambda$ (small excitonic density) is the natural expansion parameter. Since $\rho^2 \approx \rho$ we can use Wick contractions in the four-operator  expectation values, and write,
\begin{equation}
\langle H \rangle = \sum \left( \varepsilon_c \rho_{cc} + \varepsilon_v \rho_{vv} \right) + \frac{1}{2}\sum W_{\alpha\beta\gamma\delta} (\rho_{\alpha\delta}\rho_{\beta\gamma} - \rho_{\alpha\gamma}\rho_{\beta\delta}).
\end{equation}
We use the shorthand $\alpha = l,\mathbf{k}$ to denote the quantum numbers associated with the particular state. $l = \lbrace c,v\rbrace$ for the conduction and valence states. 
We consider Coulomb interaction without form factor.
\begin{equation}
    W_{[l_1, k_1] [l_2, k_2] [l_3, k_3] [l_4, k_4]} = \delta_{l_1 l_4}\delta_{l_2 l_3} \delta_{k_1 + k_2, k_3 + k_4} \times [U(k_1-k_4) \text{ or } V(k_1-k_4))]
\end{equation}
where,
\begin{equation} \begin{aligned}
\quad &\text{if } l_1 = l_2, \text{we use } V(k_1 - k_4), \quad V = \frac{2\pi e^2}{\epsilon q} \\
&\text{if } l_1 \neq l_2, \text{we use } U(k_1 - k_4),  \quad U = \frac{2\pi e^2}{\epsilon q} e^{-qd}.
 \end{aligned} \end{equation}

Then the kinetic energy terms can be written as,
\begin{equation} \begin{aligned}
\sum_c \varepsilon_c (\Lambda^\dagger \Lambda)_{cc} + \sum_v \varepsilon_v (\delta_{vv} - (\Lambda\Lambda^\dagger)_{vv}) &= \sum_{vc} \varepsilon_c \Lambda_{cv}^\dagger \Lambda_{vc} + \sum_v \varepsilon_v - \sum_{vc} \varepsilon_v \Lambda_{vc} \Lambda_{cv}^\dagger \\
&= \text{constant} + \sum_{cv} (\varepsilon_c - \varepsilon_v) |\Lambda_{vc}|^2
 \end{aligned} \end{equation}
This term is quadratic in $\Lambda$.
In the ordinary excitonic Hamiltonian, there is another quadratic term (interaction between electron and hole), which arises from the microscopic Fock term, to be discussed later.

\subsection*{Hartree terms}
We express the Hartree term in terms of $\Lambda$,
\begin{equation} \begin{aligned}
\frac{1}{2} W_{\alpha\beta\gamma\delta} \, \rho_{\alpha\delta} \, \rho_{\beta\gamma} &= \frac{1}{2} W_{\text{K}\, c_1,\, K\, c_2,\, K\, c_3,\, K\, c_4} \, \rho_{c_1 c_4} \, \rho_{c_2 c_3} + \frac{1}{2} W_{K\, c_1,\, K'\, v_2,\, K'\, v_3,\, K\, c_4} \, \rho_{c_1 c_4} \, \rho_{v_2 v_3} \\
&\quad + \frac{1}{2} W_{K'\, v_1,\, K'\, v_2,\, K'\, v_3,\, K'\, v_4} \, \rho_{v_1 v_4} \, \rho_{v_2 v_3} + \frac{1}{2} W_{K'\, v_1,\, K\, c_2,\, K\, c_3,\, K'\, v_4} \, \rho_{v_1 v_4} \, \rho_{c_2 c_3} \\[1.5ex]
&= \frac{1}{2} W_{c_1 c_2 c_3 c_4} (\Lambda^\dagger \Lambda)_{c_1 c_4} (\Lambda^\dagger \Lambda)_{c_2 c_3} + \frac{1}{2} W_{v_1 v_2 v_3 v_4} \left(\delta_{v_1 v_4} - (\Lambda\Lambda^\dagger)_{v_1 v_4}\right) \left(\delta_{v_2 v_3} - (\Lambda\Lambda^\dagger)_{v_2 v_3}\right) \\
&\quad + \frac{1}{2} W_{c_1 v_2 v_3 c_4} (\Lambda^\dagger \Lambda)_{c_1 c_4} \left(\delta_{v_2 v_3} - (\Lambda\Lambda^\dagger)_{v_2 v_3}\right) + \frac{1}{2} W_{v_1 c_2 c_3 v_4} \left(\delta_{v_1 v_4} - (\Lambda\Lambda^\dagger)_{v_1 v_4}\right) (\Lambda^\dagger \Lambda)_{c_2 c_3}
 \end{aligned} \end{equation}
The term with $\delta_{v_2 v_3}$ gives self-energy corrections which are absorbed into $\varepsilon_v$.

\subsection*{Fock terms}
Fock term can be rewritten in terms of $\Lambda$ as,
\begin{equation} \begin{aligned}
-\left[ \vphantom{\frac{1}{2}} \right.
&\quad \frac{1}{2} W_{c_1 c_2 c_3 c_4} (\Lambda^\dagger \Lambda)_{c_1 c_3} (\Lambda^\dagger \Lambda)_{c_2 c_4} \\
&+ \frac{1}{2} W_{v_1 v_2 v_3 v_4} \left(\delta_{v_1 v_3} - (\Lambda \Lambda^\dagger)_{v_1 v_3}\right) \left(\delta_{v_2 v_4} - (\Lambda \Lambda^\dagger)_{v_2 v_4}\right) \\
&+ \frac{1}{2} W_{c_1 v_2 v_3 c_4} \left(\Lambda^\dagger - \frac{1}{2} \Lambda^\dagger \Lambda \Lambda^\dagger\right)_{c_1 v_3} \left(\Lambda - \frac{1}{2} \Lambda \Lambda^\dagger \Lambda\right)_{v_2 c_4} \\
&+ \frac{1}{2} W_{v_1 c_2 c_3 v_4} \left(\Lambda - \frac{1}{2} \Lambda \Lambda^\dagger \Lambda\right)_{v_1 c_3} \left(\Lambda^\dagger - \frac{1}{2} \Lambda^\dagger \Lambda \Lambda^\dagger\right)_{c_2 v_4} \left. \vphantom{\frac{1}{2}} \right]
 \end{aligned} \end{equation}
From this, we isolate a term quadratic in $\Lambda$ and add it to the kinetic energy, to get the {total quadratic bosonic term}.

The total energy is $E = {\rm{constants}} + H_{\rm quad} + H_H^{(4)} + H_X^{(4)}$ where each of the terms are defined below.

The {total quadratic bosonic term} takes the form,
\begin{equation}
H_{\rm quad} = \sum (\varepsilon_c - \varepsilon_v) |\Lambda_{vc}|^2 - \frac{1}{2} W_{c_1 v_2 v_3 c_4} \Lambda_{c_4 v_2} \Lambda_{v_3 c_1}^\dagger - \frac{1}{2} W_{v_1 c_2 c_3 v_4} \Lambda_{v_1 c_3} \Lambda_{c_2 v_4}^\dagger
\end{equation}
which is the Bethe-Salpeter equation in disguise.

From the fermionic Hartree term, we get a \textbf{bosonic Hartree-like term},
\begin{equation}
H_H^{(4)} = \frac{1}{2} W_{c_1 c_2 c_3 c_4} (\Lambda^\dagger\Lambda)_{c_4 c_1} (\Lambda^\dagger\Lambda)_{c_3 c_2} + \frac{1}{2} W_{v_1 v_2 v_3 v_4} (\Lambda\Lambda^\dagger)_{v_4 v_1} (\Lambda\Lambda^\dagger)_{v_3 v_2} - W_{c_1 v_1 v_2 c_2} (\Lambda^\dagger\Lambda)_{c_2 c_1} (\Lambda\Lambda^\dagger)_{v_1 v_2}
\end{equation}

From the fermionic Fock terms quartic in $\Lambda$ we get the \textbf{bosonic Fock term},
\begin{equation} \begin{aligned}
H_X^{(4)} = &+ {\frac{1}{2} W_{c_1 v_1 v_2 c_2} \left( \Lambda_{c_1 v_2}^\dagger (\Lambda \Lambda^\dagger \Lambda)_{v_1 c_2} + \Lambda_{v_1 c_2} (\Lambda^\dagger \Lambda \Lambda^\dagger)_{c_1 v_2} \right)}\\
&-{\left[\frac{1}{2} W_{c_1 c_2 c_3 c_4} (\Lambda^\dagger \Lambda)_{c_1 c_3} (\Lambda^\dagger \Lambda)_{c_2 c_4} +\frac{1}{2} W_{v_1 v_2 v_3 v_4} (\Lambda \Lambda^\dagger)_{v_1 v_3} (\Lambda \Lambda^\dagger)_{v_2 v_4}\right]}.
 \end{aligned} \end{equation}

We denote the first term as `Type-I' Fock term and the second term as `Type-II'.

The valley polarization is determined by the bosonic Fock terms.
To get the values of $\Lambda$, we solve the quadratic eigenvalue equation (Bethe-Salpeter equation),
\begin{equation}
\sum_{k'} \left[(\varepsilon_{ck} - \varepsilon_{vk}) \delta_{kk'} - \frac{1}{A} U(k - k')\right] \Lambda_{vk', ck'} = -E_{\rm bound} \Lambda_{vk, ck}.
\end{equation}
where $\Lambda_{vk, ck} = \frac{1}{\sqrt{A}} f_k B_{vc}$, with normalization $\frac{1}{A}\sum |f_k|^2 = 1$.

From this, we obtain,
\begin{equation} \begin{aligned}
(\Lambda^\dagger \Lambda)_{ck, c'k'} &= (\Lambda^\dagger)_{ck, vk''} \Lambda_{vk'', c'k'} = \frac{1}{A} \delta_{kk'} |f_k|^2 (B^\dagger B)_{cc'} \\
(\Lambda \Lambda^\dagger)_{vk, v'k'} &= \frac{1}{A} \delta_{kk'} |f_k|^2 (B B^\dagger)_{vv'}
 \end{aligned} \end{equation}
with $\text{Tr}(B^\dagger B) = \text{Tr}(B B^\dagger) = N_{\text{ex}}(k)$.

The $B$ matrices keep track of the type or flavor (inter-valley/intra-valley) of excitons. We ignore spin, for simplicity. 

\subsection*{Fock terms of Type I and Type II}
The total Type-I contribution is $+E_X^{(1)}$ with,
\begin{equation} \begin{aligned}
E_X^{(1)}&= \frac{1}{2} W_{c_1 v_1 v_2 c_2} \Lambda_{c_1 v_2}^\dagger (\Lambda \Lambda^\dagger \Lambda)_{v_1 c_2} + {h.c.}\\
&= \frac{1}{2A} \sum_{k_1, k_2} U(k_1-k_2) f^*_{k_1} f^*_{k_2} |f_{k_2}|^2 \text{tr}[(B B^\dagger)^2] + h.c. \\
&= g_x^{(1)} \text{Tr}[(B B^\dagger)^2].
 \end{aligned} \end{equation}

The Type-II term is $-E_X^{(2)}$ with,
\begin{equation} \begin{aligned}
E_X^{(2)} &= \frac{1}{A} \sum_{k_1, k_2} V(k_1 - k_2) |f(k_1)|^2 |f(k_2)|^2 \text{Tr}\left[(B^\dagger B)^2\right] \\
&= g_X^{(2)} \text{Tr}\left[(B^\dagger B)^2\right].
 \end{aligned} \end{equation}

Now,
\begin{equation}
\text{Tr}[(B B^\dagger)^2] = \text{Tr}\left[(B^\dagger B)^2\right] = n_1^2 + n_2^2
\end{equation}
where $n_1$ and $n_2$ are excitonic densities in valley $K$ and $K'$, respectively.

Therefore the total Fock energy is,
\begin{equation}
    E_X = E_X^{(1)} - E_X^{(2)} =  (g_X^{(1)} - g_X^{(2)}) (n_1^2 + n_2^2).
\end{equation}

In comparison, the Hartree term contains terms like $(\text{Tr}[B^\dagger B])^2 = (n_1 + n_2)^2 = N^2$.

\subsection*{Free Energy Expression}
Substituting the values of $\Lambda$ from the solution of Bethe-Salpeter equation in the expressions of different terms in the bosonic interaction Hamiltonian, we obtain an expression for the free energy (this is really the expression of energy at zero temperature obtained from the mean-field ansatz),
\begin{equation}
\begin{aligned}
F &= -a (n_1 + n_2) + g_H (n_1 + n_2)^2 + \left(g_X^{(1)} - g_X^{(2)}\right)(n_1^2 + n_2^2) \\
&= -a (n_1 + n_2) + \left(g_H + \frac{g_X}{2}\right) (n_1 + n_2)^2 + \frac{g_X}{2}(n_1 - n_2)^2\\
&= - a N + \left(g_H + \frac{g_X}{2}\right) N^2 + \frac{g_X}{2} \Delta^2
\end{aligned}
\end{equation}
where $N = n_1 + n_2$ and $\Delta = (n_1 - n_2)$, and $a$ is a constant.

\textbf{Clearly, if $g_X < 0$, then one valley will be spontaneously polarized}, sending all the electrons to a single valley, whereas if $g_X >0$ then the excitons will remain unpolarized ($\Delta = (n_1 - n_2) = 0$).

The condition for valley-polarization has \textbf{no threshold as it is only quadratic in $\Delta$}.
At large excitonic densities there will be $\mathcal{O}(\Delta^4)$ terms which will set the upper limit of valley polarization. But these arise from terms of order $\Lambda^6$ or $\Lambda^8$, whereas this theory is only self-consistent upto $\Lambda^5$.

\textit{Specific example of valley polarization}:
Consider the \textit{Mexican hat dispersion model} where all the wavefunction is peaked near $|\mathbf k| \approx k_0$, such that $f_k$ has nearly constant magnitude there. 
Then 
\begin{equation}
   g_X =  g_X^{(1)} - g_X^{(2)}\approx \sum_{k_1, k_2} |f|^4 [U(k_1 - k_2) - V(k_1 - k_2)] < 0,
\end{equation}
leading to valley polarization.

\end{document}